\documentclass[
reprint,
superscriptaddress,
amsmath,amssymb,
aps,
pra,
floatfix,
]{revtex4-2}
\usepackage{graphicx}
\usepackage{amsmath}
\usepackage{amssymb}
\usepackage{mathtools}            % for 'dcases*' env.
\usepackage[dvipsnames]{xcolor}
\usepackage{siunitx}
\usepackage{subcaption}
\usepackage{dcolumn}
\usepackage{bm}
\usepackage{soul}
\usepackage{orcidlink}

\begin{document}

\title{Using coherent feedback for a periodic clock}

\author{Stefan Zeppetzauer\,\orcidlink{0000-0003-4546-408X}}
\email[To whom correspondence should be addressed: ]{s.zeppetzauer@uq.edu.au}
\affiliation{ARC Centre of Excellence for Engineered Quantum Systems, School of Mathematics and Physics, The University of Queensland, St Lucia, Australia.}
\author{Leonardo Assis Morais\,\orcidlink{0000-0003-1617-979X}}
\affiliation{ARC Centre of Excellence for Engineered Quantum Systems, School of Mathematics and Physics, The University of Queensland, St Lucia, Australia.}
\affiliation{Instituto Federal Fluminense, Campus Bom Jesus do Itabapoana, Bom Jesus do Itabapoana, Rio de Janeiro, Brazil}
\author{Xin He\,\orcidlink{0000-0002-2913-6388}}
\affiliation{ARC Centre of Excellence for Engineered Quantum Systems, School of Mathematics and Physics, The University of Queensland, St Lucia, Australia.}
\author{Gerard Milburn}
\affiliation{ARC Centre of Excellence for Engineered Quantum Systems, School of Mathematics and Physics, The University of Queensland, St Lucia, Australia.}
\author{Arkady Fedorov}
\affiliation{ARC Centre of Excellence for Engineered Quantum Systems, School of Mathematics and Physics, The University of Queensland, St Lucia, Australia.}

\date{\today}

\begin{abstract}
A driven linear oscillator and a feedback mechanism are two necessary elements of any classical periodic clock.  Here, we introduce a novel, fully quantum clock using a driven oscillator in the quantum regime and coherent quantum feedback. We show that if we treat the model semiclassically, this system supports limit cycles, or self-sustained oscillations, as needed for a periodic clock. We then analyse the noise of the system quantum mechanically and prove that the accuracy of this clock is higher compared to the clock implemented with the classical measurement feedback. We experimentally implement the model using two superconducting cavities with incorporated Josephson junctions and microwave circulators for the realisation of the quantum feedback. We confirm the appearance of the limit cycle and study the clock accuracy both in frequency and time domains. Under specific conditions of noisy driving, we observe that the clock oscillations are more coherent than the drive, pointing towards the implementation of a quantum autonomous clock.
\end{abstract}

\maketitle

\section{Introduction}
\label{sec:intro}
%###############################

%\textcolor{red}{``Definition" of a clock. How a clock works? Which are the necessary conditions in order to build a clock? How the measurement-based feedback works?}
%\textcolor{red}{Test the limits of clock performance due to quantum noise. Clock at 0 temperature, only noise mechanism present are quantum fluctuations. }
%\textcolor{red}{Coherent feedback as a strategy to reduce the noise in the clock operation.}

% importance of clocks
%\stefan{Expand or delete}
%Precise and stable clock technology underwrites economic growth and security. Synchronising telecommunication networks and power grids and time-stamping of transactions and data transmissions is critical national infrastruture. Precise timekeeping is crucial for satellite navigation and communication systems. Quite apart from huge public investments in clock design, private investment is rising rapidly. The size of the market for compact atomic clocks is expected to exceed around US$1$ billion by 2033~\cite{fmi}. \arkady{Some text about quantum regime and why it is expected that clocks will pushed into this regime.}

A clock is a driven, stable oscillator (limit cycle) with a counter~\cite{levine_introduction_1999}. Clocks make use of feedback to regulate the period of the oscillation~\cite{milburn_thermodynamics_2020} (see Appendix~\ref{appendix1} for more details on the role of limit cycles in clock design). 
%A suitable physical quantity must be continuously monitored to enable the implementation of the counter.  
In classical physics, all feedback is equivalent to measurement-based feedback. In a quantum clock oscillator, feedback based on continuous measurement necessarily adds noise as the measurement currents carry quantum measurement noise and this drives noise in the clock oscillator. However, there is a distinctly quantum form of feedback, called coherent feedback,  that makes use of circulators to break time-reversal invariance. This form of feedback does not use measurement currents but rather uses unidirectional propagating quantum fields directly\cite{wiseman_all-optical_1994, wiseman_quantum_2009}. A quantum clock built using coherent feedback must necessarily be less noisy than a clock that uses measurement feedback control.  

Here, we propose and demonstrate a novel quantum clock using coherent feedback in a superconducting circuit. Since the feedback does not involve readout, there is no quantum measurement noise imposed on the feedback signal, allowing us to investigate and characterise the inherent quantum noise of the system and its effects on the clock properties. Our quantum clock is composed of two high-Q cavities coupled by coherent feedback \cite{wiseman_all-optical_1994}. A semiclassical analysis shows that limit cycles are possible and can therefore be used to implement stable clock oscillations. This analysis does not take into account noise. In order to find the precision of the clock, both quantum and classical noise need to be included. At low temperature, the quantum noise dominates, and  in this limit, the precision of the clock can be bounded by the recently developed thermodynamic uncertainty relations~\cite{kewming2023passage}.

% From a thermodynamic point of view, clocks are dissipative systems which uses the increase in entropy to measure the passage of time \cite{milburn_thermodynamics_2020}. For periodic clocks, their operation is related to a presence of a stable limit cycle: an isolated periodic trajectory in phase space. Limit cycles are self-sustained oscillations that arises when the energy dissipated over a cycle is matched by the energy delivered to the system through the driving mechanism. An example is the pendulum clock , where the kicks imposed in each cycle provides the required energy for the clock to keep constantly ticking when the dissipative effects are taken into account.
% \vskip 0.5 truecm
% {\textcolor{magenta}{GJM: Add appendix to discuss the role of limit cycles in  clocks and Wald distribution. }}
% \vskip 0.5 truecm

This paper is structured as follows. Section~\ref{sec:model} presents the proposed model for a quantum clock using coherent feedback and derives the master equation that describes its dynamics using the SLH framework. The semiclassical equations of motion are derived from the master equation, and the limits where this approximation holds are discussed. A linear stability analysis of the fixed points is performed, showing that a supercritical Hopf bifurcation appears in the system. The master equation for a measurement-based feedback clock is also derived to demonstrate the equivalency to the coherent feedback clock. Section~\ref{sec: noise analysis} compares how noise affects both measurement-based and coherent feedback clocks, theoretically demonstrating the advantage of coherent feedback for clock accuracy. Section~\ref{sec: experimental realisation} details the experimental realisation of the model using superconducting microwave resonators and Josephson junctions. The results of the measurements are presented in Section~\ref{sec: results}, demonstrating the emergence of limit cycles and studying their properties and the resulting clock accuracy as a function of external parameters. The results are summarised, and the outlook is presented in Section~\ref{sec: conclusion}.

%In Section 5, we solve the time evolution for the quadratures of the clock cavity, showing numerical evidence of the presence of limit cycle. Trajectories in the phase space are also shown. The clock resolution is extracted from the limit cycle periods, and we show how it varies when changing some experimental parameters. To conclude, we briefly discuss our next steps, which includes an experimental realisation of our proposal using superconducting circuits and a full quantum analysis using continuous measurements.

%Independently of the physical system chosen to work as the clock, we need to connect this system to a counter. Only then we will be able to register the number of counts, allowing us to measure the passage of time. 

% In section 4, we show our clock signal: the quantum trajectories solutions obtained for the stochastic master equation. From these trajectories we can measure the clock ticks and evaluate clock accuracy and resolution. In section 5, we quantify the amount of energy spent in one clock oscillation, and relates the entropy generated in this process with the accuracy of the clock. In section 6, we present a possible implementation of our model using superconducting circuits, along with further directions to explore coherent feedback in the context of precise time measurements.

%###############################

\section{Model}
\label{sec:model}

%###############################

%\textcolor{red}{Explain the model used with two optical cavities and coherent feedback.}

%\textcolor{red}{Add master equation, explicitly showing Hamiltonian and decay terms.}

%\textcolor{red}{Make an schematic representation of the system.}
% The objective of the controller cavity is create oscillations in a regular fashion that we can use as the ticks for our clock.

\begin{figure*}
    \includegraphics[width=\textwidth]{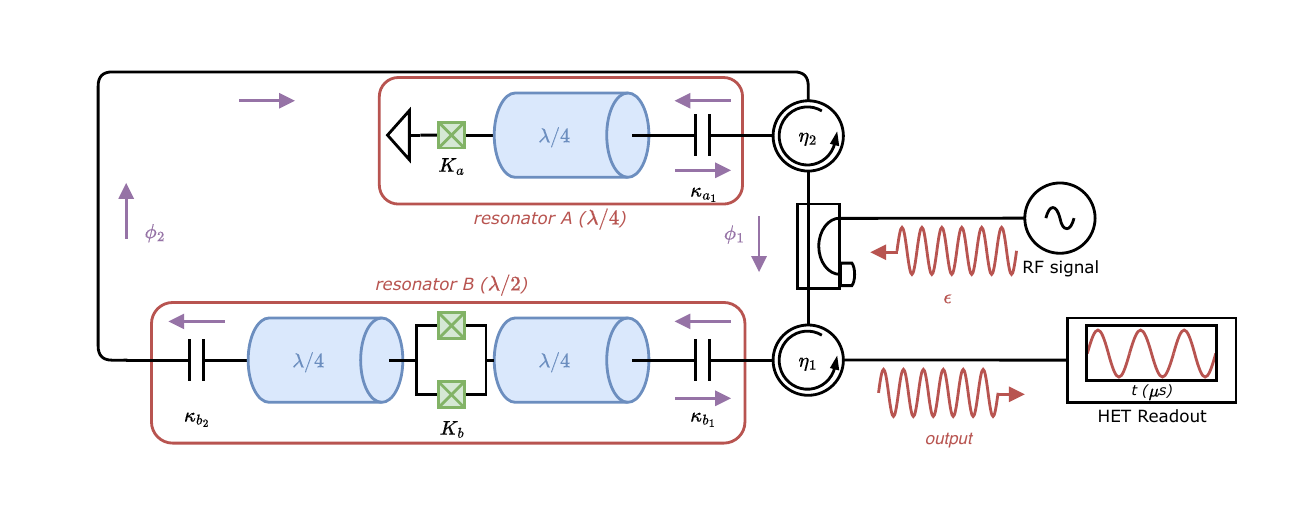}
    \caption{\label{fig: fb loop schematic} Coherent feedback clock schematic: The incoming drive is injected into the circuit by a directional coupler. The clock resonator of length $\lambda/2$ is coupled to a transmission line on both ends via capacitors with coupling rates $\kappa_{b_1}$ and $\kappa_{b_2}$. A SQUID is embedded in the centre of the resonator, consisting of a loop of Josephson junctions, depicted as a green box with a cross to tune its frequency by an external flux threaded through the SQUID loop. Microwaves reflected from the input port are fed into the readout chain for heterodyne (HET) readout by the first circulator. Signal leaking through another end is directed to the feedback resonator by the second circulator. The feedback resonator of length $\lambda/4$ is shorted to ground via a Josephson junction which generates the Kerr nonlinearity $K_a$ on one end and capacitively coupled to a transmission line with rate $\kappa_{a_{1}}$. Incoming signals are reflected and eventually fed back to the clock resonator.}\label{fig: 1 scheme}
\end{figure*}

Our clock scheme is shown in Fig.~\ref{fig: fb loop schematic}. It consists of a superconducting microwave circuit composed of two coplanar waveguide resonators: a quarter-wavelength resonator (A) and a half-wavelength (B). They are connected by microwave transmission lines via coupling capacitors.  Resonator B is coherently driven by an external electromagnetic field of amplitude $\epsilon$ and frequency $\omega_\epsilon$. With respect to the driving field, both resonators are detuned by 
\begin{equation}
\label{eq: cav detuning}\Delta_i = \omega_i - \omega_\epsilon,
\end{equation}
where $\omega_i$ is the resonance frequency for resonator $i$, with $i = a, b$. Both resonators incorporate Josephson junctions providing a Kerr-type nonlinearity of strength $K_i$. The total decay rates of the resonators, $\kappa_i$, consist of the coupling rates for each side of the resonators and the internal losses.

To study the dynamics of our clock system, we use the master equation under Born-Markov approximations \cite{gardiner_quantum_2004}:
\begin{equation}
    \label{eq:master_equation}
    \dot\rho = -\frac{i}{\hslash} [\hat{H}, \rho] + \sum_i \mathcal{D}[\hat{L}_i]\rho,
\end{equation}
where $\rho$ is the composite quantum state that describes the intracavity field for both resonators, $H$ is the system Hamiltonian, and $L_i$ are the collapse operators. The superoperator $\mathcal{D}$ is defined as 
\begin{equation}
    \mathcal{D}[\hat{L}_i]\rho = \hat{L}_i^\dagger \rho \hat{L}_i - \frac{1}{2}(\hat{L}_i^\dagger \hat{L}_i \rho + \rho \hat{L}_i^\dagger \hat{L}_i ).
\end{equation}

To obtain the Hamiltonian $\hat{H}$ and the jump operators $\hat{L}_i$ for our coherent clock, we use the SLH framework~\cite{gough_series_2009, gough_quantum_2009}, which provides a set of algebraical rules to model complex quantum networks from basic components. Details about the decomposition of our system in terms of SLH components are discussed in Appendix \ref{appendix2}. The resulting Hamiltonian that describes our model is given by (note that $\hslash = 1$)
% Lossy:
% \begin{equation}
%     \begin{split}
%     \hat{H} & = K_{a} \hat{a}^{\dagger2} \hat{a}^2 + \Delta_{a} \hat{a}^{\dagger} \hat{a} + K_b \hat{b}^{\dagger2} \hat{b}^2 + \Delta'_{b} \hat{b}^{\dagger} \hat{b}  \\
%     & + \frac{i}{2}\sqrt{\kappa_{a_{1}}}\left(\sqrt{\kappa_{b_{1}}}\sqrt{1-\eta^2}e^{-i\phi} - \sqrt{\kappa_{b_{2}}} \right)\hat{a}^\dagger \hat{b} +c.c. \\
%     &-i\sqrt{\kappa_{b_{1}}}\sqrt{1-\eta^2}\epsilon \hat{b}^\dagger + c.c.         
%     \end{split},
%     \label{eq:clock_hamiltonian}
% \end{equation}
% No loss:
\begin{align}
    \hat{H} & = K_{a} \hat{a}^{\dagger2} \hat{a}^2 + \Delta_{a} \hat{a}^{\dagger} \hat{a} + K_b \hat{b}^{\dagger2} \hat{b}^2 + \Delta'_{b} \hat{b}^{\dagger} \hat{b}  \nonumber\\
    & + \frac{i}{2}\sqrt{\kappa_{a_{1}}}\left[\sqrt{\kappa_{b_{1}}\left(1-\eta_1^2\right)}e^{-i\phi_1} - \sqrt{\kappa_{b_{2}}(1-\eta_2^2)}e^{i \phi_2} \right]\hat{a}^\dagger \hat{b} \nonumber\\
    & -i\sqrt{\kappa_{b_{1}}}\epsilon \hat{b}^\dagger + c.c.,
    \label{eq:clock_hamiltonian}
\end{align}
with
% Lossy
% \begin{equation}
%     \label{eq:effective_detuning_b}
%     \Delta'_b = \Delta_b + \stefan{\sqrt{1-\eta^2}} \sqrt{\kappa_{b_{1}}\kappa_{b_{2}}} \sin{(\phi)},
% \end{equation}. 
% No loss
\begin{equation}
    \label{eq:effective_detuning_b}
    \Delta'_b = \Delta_b + \sqrt{\kappa_{b_{1}}\kappa_{b_{2}}}\sqrt{(1-\eta_1^2)(1-\eta_2^2)} \sin{(\phi_1+\phi_2)},
\end{equation}
where $\hat i(\hat i^\dagger)$ is the creation (annihilation) operator, $K_{a,b}$ are the Kerr nonlinearity coefficients, $\kappa_{a_{1},b_{1},b_{2}}$ are the coupling rates of the resonator ports and $\epsilon$ is the amplitude of the driving field.
While the loss mechanisms from elements between the resonators such as the circulators are typically small at operating temperatures, they still influence the parameter regime where limit cycles occur and should not be neglected. We account for losses in the circulators by the insertion loss parameter $0\leq \eta_{i} \leq 1$ with $i=1,2$. It is defined as the amplitude ratio of the lost (absorbed or reflected) to incident signal in the forward direction. The phases, $\phi_{i}$, are the acquired phases for the electromagnetic field while it travels between cavities, and $\epsilon$ is the amplitude of the driving field.  
The quadratic terms describe the Kerr-type nonlinearity introduced in each cavity by the Josephson junction. Note that resonator B can be, in principle, fully linear. Our reason for introducing a tunable nonlinearity, in form of a superconducting quantum interference device (SQUID), is to be able to modulate the detuning to reach the necessary conditions for the emergence of a limit cycle.
The collapse operators for the superconducting clock are
% \vskip 1 truecm
\begin{widetext}
   \begin{equation}
    \bm{\hat{L}}=\begin{pmatrix} {\rm  - \eta_1 \sqrt{\kappa_{a_{1}}} e^{ i \phi_1} \hat{a}} -  \eta_1 \sqrt{\kappa_{b_{2}}} \sqrt{1-\eta_2^2} e^{i \left(\phi + \phi_{2}\right)} \hat{b} \\  \sqrt{\kappa_{a_{\mathrm{int}}}} \hat{a} \\ {\rm  \sqrt{\kappa_{a_{1}}} \sqrt{1-\eta_1^2} e^{i \phi_1} \hat{a}} +  \left(\sqrt{\kappa_{b_{1}}} + \sqrt{\kappa_{b_{2}}} \sqrt{(1-\eta_1^2)(1-\eta_2^2)} e^{i \left(\phi_1 + \phi_{2}\right)}\right) \hat{b} \\  - \eta_{2} \sqrt{\kappa_{b_{2}}} e^{i \phi_{2}} \hat{b} \\  \sqrt{\kappa_{b_{\mathrm{int}}}} \hat{b}\end{pmatrix}
        \label{eq:collapse_op},
    \end{equation}
\end{widetext}
% \begin{equation}
%     \bm{\hat{L}}=\begin{pmatrix}
%     \sqrt{\kappa_{aint}} \hat{a} \\   
%     \sqrt{\kappa_{a_{1}}} \hat{a} +  \left(\sqrt{\kappa_{b_{1}}} + \sqrt{\kappa_{b_{2}}} \right) \hat{b} \\  
%     \sqrt{\kappa_{b_{\mathrm{int}}}} \hat{b}\end{pmatrix}
%     \label{eq:collapse_op}
% \end{equation}
where the first entry describes the feedback between the cavities and the first and third entry account for the internal losses of the resonators given by the rates $\kappa_{i_{\mathrm{int}}}$ with $i=a,b$. 
It is worth noting that we neglect additional loss mechanism and imperfections such as reflections and circulator isolation, which should be taken into account if a more precise prediction is required. Since we are only interested to find an approximate parameter regime, we do not include these factors here.
% Old version:
% \begin{equation}
%     \hat{L} = \begin{pmatrix}\eta^{2} \sqrt{\kappa_{a_{1}}} \hat{a} + \eta \left(\eta e^{ i \phi} \sqrt{\kappa_{b_{2}}}  + \sqrt{\kappa_{b_{1}}}\right) \hat{b} \\
%     \sqrt{1- \eta^{2}} \left[\eta \sqrt{\kappa_{a_{1}}} \hat{a} + \left(\eta e^{i \phi} \sqrt{\kappa_{b_{2}}}  + \sqrt{\kappa_{b_{1}}}\right) \hat{b} \right] \\
%     \sqrt{1- \eta^{2}} \left[ \sqrt{\kappa_{a_{1}}} \hat{a} + e^{i \phi} \sqrt{\kappa_{b_{2}}}    \hat{b}\right]\end{pmatrix}
%     \label{eq:collapse_op}
% \end{equation}
%The SLH acronym indicates the three elements required to describe a circuit component: the scattering matrix $S$, the collapse operators $L_i$, and the Hamiltonian $H$. For a comprehensive review on the SLH framework, the reader is referred to Ref. \cite{combes_slh_2017}. Each component in a quantum network can be described by a SLH triple: 
% \begin{equation}
%     G_i = (\mathbf{S_i}, \mathbf{L_i}, H_i),
% \end{equation}
% where $\mathbf{S_i}$ is the scattering matrix, $\mathbf{L_i}$ is a vector containing all jump operators, and $H_i$ is the Hamiltonian for the element $i$ in the network. Below we present the two composition rules we will use for the master equation derivation for the coherent clock, as well as the SLH triples for the basic components of our system. %By applying these rules together, we build our SLH triple.

\subsection{Obtaining semiclassical equations}

Using the Hamiltonian and collapse operators, we can calculate the master equation from which the expectation values for the annihilation operators for both cavities can be obtained by
\begin{equation}
    \label{eq:ab_evo}
    \frac{\mathrm{d}}{\mathrm{dt}} \left\langle \hat{a} \right\rangle  = \mathrm{Tr}[ \hat{a}\dot{\rho}], \quad \frac{\mathrm{d}}{\mathrm{dt}} \langle \hat{b} \rangle = \mathrm{Tr}[\hat{b}\dot{\rho}].
\end{equation}
From this, we get:
\begin{align}
        \frac{\mathrm{d}}{\mathrm{dt}} \left\langle \hat{a} \right\rangle = &-\left( i \Delta_a + \frac{\kappa_{a}}{2} \right)\langle \hat{a} \rangle - 2 i K_a \langle \hat{a}^\dagger \hat{a}^2 \rangle \nonumber\\
        & -  \sqrt{1-\eta_2^2} e^{i\phi_2} \sqrt{\kappa_{a_{1}} \kappa_{b_{2}}}  \langle \hat{b} \rangle,
    \label{eq:evolution_a}
\end{align}
\begin{align}
    \label{eq:evolution_b}
    \frac{\mathrm{d}}{\mathrm{dt}} \langle \hat{b} \rangle= & - \left( i\Delta'_b  + \frac{\kappa_b}{2}  \right)  \langle \hat{b} \rangle - \sqrt{1-\eta_1^2} e^{i\phi_1}\sqrt{\kappa_{a_{1}} \kappa_{b_{1}}} \langle \hat{a} \rangle \nonumber\\
    & -\sqrt{1-\eta_1^2}\sqrt{\kappa_{b_{1}}} \epsilon,
\end{align}
% {\textcolor{red}{What is $a_m$? I think we should set $\phi=0$ as it is not used in the simulations. Is there any way to measure it? }} \stefan{I used it in the simulations as a free parameter to check the properties of the LCs at different values. But we can't measure it directly, I think, as it depends on the exact path length between the resonators. However, the limit cycle properties are highly dependent on this phase, so I'm not sure if we can just assume it to be zero.}
where we define the total loss rates
\begin{equation}
    \kappa_{a}=\kappa_{a_{1}}+\kappa_{a_{\mathrm{int}}}   
\end{equation} 
and
\begin{align}
    \label{eq:effective_decay_rate_b}
    \kappa_b =& \kappa_{b_{1}} +\kappa_{b_{2}} + \kappa_{b_{\mathrm{int}}} \nonumber \\
    &+ 2  \sqrt{\kappa_{b_{1}}\kappa_{b_{2}}} \sqrt{(1-\eta_1^2)(1-\eta_2^2)} \cos{(\phi_1+\phi_2)}.
\end{align}
The semiclassical approximation is obtained by associating $\alpha$ and $\beta$, the classical complex amplitudes of the intracavity electromagnetic field, with the average values $\langle a \rangle$ and  $\langle b \rangle$ obtained from the master equation, respectively. Additionally, we assume that the intracavity states remain approximately coherent, i.e.,
\begin{equation}
    \langle a^\dagger a^2 \rangle  \approx \alpha |\alpha|^2.
\end{equation}
The semiclassical equations of motion for our coherent feedback clock are:
\begin{align}
     \label{eq:classical_a}
    \dot\alpha & =  -\left( i \Delta_a + \frac{\kappa_a}{2} \right) \alpha - 2 i K_a |\alpha|^2 \alpha - g_a e^{i\phi_2}\beta \\ 
   \dot\beta & =- \left( i\Delta'_b  + \frac{\kappa_b}{2} \right)  \beta - 2 i K_b |\beta|^2 \beta-g_b e^{i\phi_1} \alpha - \bar{\epsilon}, \label{eq:classical_b}  
\end{align}
%\stefan{\sqrt{1-\eta^2}e^{i \phi}}
where $g_a=\sqrt{1-\eta_2^2}  \sqrt{\kappa_{a_{1}} \kappa_{b_{2}}}$ and $g_b=\sqrt{1-\eta_1^2}  \sqrt{\kappa_{a_{1}}\kappa_{b_{1}}}$ are coupling constants arising from coherent feedback. The effective driving field of cavity B is $\bar{\epsilon}=\sqrt{1-\eta_1^2}\sqrt{\kappa_{b_{1}}} \epsilon$. 

Since these two equations have complex solutions, we define
\begin{equation} \label{comples amplitudes}
    \alpha(t) \coloneq x_a(t) + i y_a(t), \quad \beta(t) \coloneq x_b(t) + i y_b(t).
\end{equation}
The dynamics of the dynamical system can then be analysed using four equations in terms of the field quadratures $x_j, y_j$, with $j = a, b$ as follows
\begin{align}
\label{eq:classical_xa}
    \dot{x}_a =& 2 K_a y_a^3 + 2 K_a x_a^2 y_a + \Delta_a y_a - \frac{\kappa_a}{2} x_a \nonumber\\
    &- g_a (\cos(\phi_2) x_b- \sin(\phi_2)y_b),
\end{align}
\begin{align}
\label{eq:classical_ya}
    \dot{y}_a =& -2 K_a x_a^3 - 2 K_a y_a^2 x_a - \Delta_a x_a -\frac{\kappa_a}{2} y_a \nonumber\\ 
    & - g_a (\sin(\phi_2) x_b + \cos(\phi_2)y_b) ,
\end{align}   
\begin{align}
\label{eq:classical_xb}
    \dot{x}_b =& 2 K_b y_b^3 + 2 K_b x_b^2 y_b + \Delta'_{b} y_b - \frac{\kappa_b}{2} x_b \nonumber\\ 
    &-g_b (\cos(\phi) x_a- \sin(\phi)y_a)  - \bar{\epsilon},
\end{align}       
\begin{align}
\label{eq:classical_yb}
    \dot{y}_b =&-2 K_b x_b^3 - 2 K_b y_b^2 x_b - \Delta'_b x_b - \frac{\kappa_b}{2} y_b \nonumber\\ 
    &- g_b (\sin(\phi) x_a + \cos(\phi)y_a),
\end{align}
where we assume $\epsilon \in \mathcal{R}$ without loss of generality.

The effect of the feedback is similar to parametric amplification as $x_a$ is coupled to $x_b$ and $y_a$ is coupled to $y_b$. In the case that $\kappa_{b_{1}}=\kappa_{b_{2}}=\kappa_b$ the effective gain rates for each quadrature is $\sqrt{\kappa_{a_{1}}\kappa_b}$. This would lead to an instability once this exceeds the loss rate for each cavity. However, the Kerr nonlinearity prevents this from occurring, leading to an effective gain saturation.   

% By studying the solutions of these nonlinear coupled equations, we show that the coherent clock supports limit cycles in the semiclassical approximation.

\subsection{Fixed points and stability analysis}
\label{sec:stability analysis}

\begin{figure}
    \begin{subfigure}[b]{\linewidth}
        \includegraphics[width=\linewidth]{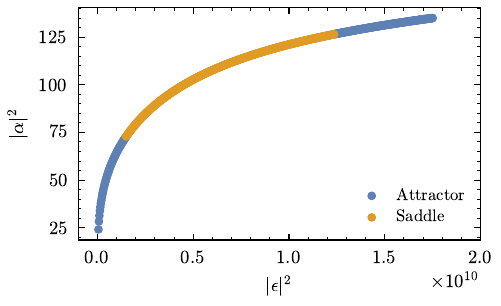}
        % \caption{\label{fig: fb loop circuit}}
        
    \end{subfigure}
    \begin{subfigure}[b]{\linewidth}
        \includegraphics[width=\linewidth]{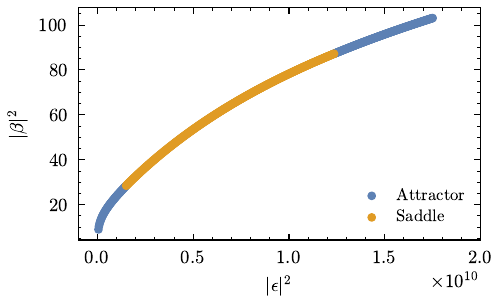}

    \end{subfigure}
    \caption{\label{fig:bistability_curves} Fixed points for the intensity of the intracavity field on resonators A and B, $|\alpha|^2$ and $|\beta|^2$, respectively, as a function of the intensity of the driving field $|\epsilon|^2$ in the semiclassical approximation. The parameters used in this simulation are: $\kappa_{b_{1}}/2\pi = \kappa_{b_{2}}/2\pi = 2.52 \times 10^6$,
    $\kappa_{b_{ \mathrm{int}}}/2\pi = 1.64 \times 10^6$,
    $\kappa_{a_{1}}/2\pi = 3.21 \times 10^6$,
    $\kappa_{a_{ \mathrm{int}}}/2\pi = 0.11 \times 10^6$,
    $\Delta_a/2\pi = \Delta_b/2\pi = 1.8 \times 10^6$, $\eta_1 =\sqrt{0.18}, \eta_2 = \sqrt{0.03}$, $\phi_1 =0.0, \phi_2 = 0.39 \pi$, $K_a/2\pi = -0.01 \times 10^6$, $K_b/2\pi = -0.03 \times 10^6 $.}
\end{figure}

Our final goal is to find a set of physical parameters that allow us to use the model proposed here as a clock. To do so, we need to evaluate the dynamics of the vector fields described by Eqs.~(\ref{eq:classical_xa}-\ref{eq:classical_yb}), analysing how their dynamics depends on the physical parameters. The first step is to find the fixed points $x_a^*, y_a^*, x_b^*, y_b^*$ for different values of the driving field $\epsilon$. We solve the equations,
\begin{equation}
    \dot{x}_a = 0, \dot{x}_b = 0, \dot{y}_a = 0, \dot{y}_b = 0,
\end{equation} 
using the Mathematica solver NSolve. Since our problem is four dimensional, we plot the intensities of the intracavity fields, 
\begin{equation}
    |\alpha|^2 = x_a^2 + y_a^2, \quad |\beta|^2 = x_b^2 + y_b^2,
\end{equation}
as a function of the intensity of the driving field, $|\epsilon|^2$. The results are shown in Fig.~\ref{fig:bistability_curves}.

% Note that, as in the case for a single cavity \cite{drummond_quantum_1980}, there is a region where three solutions are possible for a given value of $|\epsilon|^2$. 

The occurrence of limit cycles is often related to changes in the stability of fixed points. Therefore, to investigate under which conditions the limit cycles occur, we analyse the stability of the fixed points through a linear stability analysis of each fixed point. By calculating the Jacobian $\mathcal{J}$ at a fixed point $\mathbf{x}^*$, we can determine the behaviour of the vector field in the vicinity of $\mathbf{x}^*$. In our case, the $\mathcal{J}$ is a 4$\times$4 matrix with each element given by 
\begin{equation}\label{eq:jacobian}
    \mathcal{J}_{ij} = \frac{d f_i}{dz_j},
\end{equation}
where $f_i$ are the vector fields $\dot{x}_a$, $\dot{x}_b$, $\dot{y}_a$, $\dot{y}_b$ and $z_j$ is one of the four real variables $x_a, y_a, x_b, y_b$. Once $\mathcal{J}$ is found and evaluated at a given fixed point $\mathbf{x}^*$, we can find its eigenvalues and eigenvectors. From the eigenvalues $\lambda_i$, we can determine the stability of the fixed point: 
\begin{itemize}
    \item When all $\mathrm{Re}(\lambda_i) > 0,$ the fixed point is classified as a repeller or source: all nearby trajectories will exponentially diverge from the fixed point. Small perturbations will take the system away from the fixed point and the equilibrium is unstable. 
    \item When all $\mathrm{Re}(\lambda_i) < 0,$ the fixed point is classified as an attractor or drain: all nearby trajectories will exponentially converge to the fixed point. After a small perturbation, the system will go back to the fixed point and the equilibrium is stable. 
    \item A saddle point occurs when at least one of the $\mathrm{Re}(\lambda_i)$ has a different sign from the others. In this case, the stability will depend on the direction of the eigenvectors related to the eigenvalue sign: when the eigenvalue is positive (negative) the equilibrium in the respective eigenvector direction is unstable (stable).
    \item For cases where $\mathrm{Re}(\lambda_i) = 0$ the fixed point is said to be non-hyperbolic, and further analysis is required to determine the fixed point stability.
\end{itemize}

The linear stability analysis for the system parameters in Table~\ref{tab: experimental_parameters} is shown in Fig.~\ref{fig:bistability_curves}. Note that there are changes of stability around $|\epsilon|^2 \approx 0.15 \times 10^{10}$, where the fixed points lose their stability in some directions with the attractors becoming saddle points;  and $|\epsilon|^2 \approx 1.25 \times 10^{10}$, where the fixed points gain stability in some directions with the saddle points becoming attractors. In our numerical simulations we observe that, for a given set of parameters for our clock system, limit cycles occur when the $|\epsilon|^2$ is close to the values cited above. Further numerical investigations are required to classify the type of bifurcation the system undergoes. For our purposes, it is sufficient to use the stability analysis as a tool to identify the approximate parameters required to observe limit cycles.

\subsection{Semiclassical dynamical analysis}
\label{sec: semiclassical dynamical analysis}

In order to develop intuition behind the appearance of limit cycles in this system, we consider a simplified model in which all parameters are set equal except the Kerr non-linearities for which $K_a\ll K_b$. We also assume zero detuning and no driving field. As it turns out, limit cycles can arise spontaneously due to noise.

The semiclassical equation of motion for both coherent and measurement-based feedback take the form
\begin{align}
    \dot{\alpha} & = -\frac{\kappa}{2}\alpha-g\beta -2iK_a|\alpha|^2\alpha\\
     \dot{\beta} & = -\frac{\kappa}{2}\beta-g\alpha -2iK_a|\beta|^2\beta.
\end{align}
% The steady state solutions satisfy
% \begin{equation}
%     \frac{\alpha_0^2}{\beta_0^2}=\frac{1-i\mu_b|\beta_0|^2}{1-i\mu_a|\alpha_0|^2}
% \end{equation}
% where $\mu_a=4K_a/\kappa,\ \ \ \mu_a=4K_b/\kappa$. Squaring both sides
% \begin{equation}
%     |\alpha_0|^4(1+\mu_a^2|\alpha_0|^4)=|\beta_0|^4(1+\mu_b^2|\beta_0|^4) 
% \end{equation}
% One solution is $\alpha_0=\beta_0=0$. Define $X=|\alpha_0|^4,\ Y= |\beta_0|^4$. We can then write $Y=Y(X)$  which satisfies the implicit equation, 
% \begin{equation}
%     Y(X)=\frac{X(1+\mu_a^2 X)}{1+\mu_b^2Y(X)}
% \end{equation}
% Iterating this equation $N$ times we find that
% \begin{equation}
% Y_N(X)=\frac{X(1+\mu_a^2 X)}{1+N\mu_a^2\mu_b^2X}
% \end{equation}
Furthermore, we define polar coordinates as 
\begin{equation}
    \alpha= r_ae^{i\theta_a},\ \ \  \beta= r_be^{i\theta_b}.
\end{equation}
Substituting these into the equations of motion we get 
\begin{align}
    \dot{r}_a & = -\frac{\kappa}{2} r_a -gr_b \cos\phi,\\
    \dot{r}_b & = -\frac{\kappa}{2} r_b -gr_a \cos\phi,\\
    \dot{\phi} & = g\left (\frac{r_b}{r_a}+\frac{r_a}{r_b}\right )\sin\phi-2(K_ar_a^2-K_b r_b^2),
\end{align}
where $\phi=\theta_a-\theta_b$. Equations of this form have been extensively studied in the context of synchronization of limit cycles \cite{aronson1990amplitude}.  

On the synchronised limit cycle, $\dot{\phi}=0$ and we find that $r_a=r_b=r$ with 
\begin{equation}
   r^2=\frac{\sqrt{g^2-\kappa^2/4}}{|K_b-K_a|}
\end{equation}
and 
\begin{equation}
    \sin\phi =\sqrt{1-\frac{\kappa^2}{4g^2}}.
\end{equation}
The existence of the limit cycle requires $g>\kappa/2$ and $K_a\neq K_b$. To realise this system experimentally requires careful tailoring of the external and internal coupling rates and is not part of the scope of this work.

To emphasise the advantages of the coherent feedback established here, we compare it to measurement-based feedback in the following section.

\subsection{Measurement-based feedback clock}
\label{sec:MBF}
\begin{figure}
% \vskip 0.8 truecm
\includegraphics[width=\linewidth]{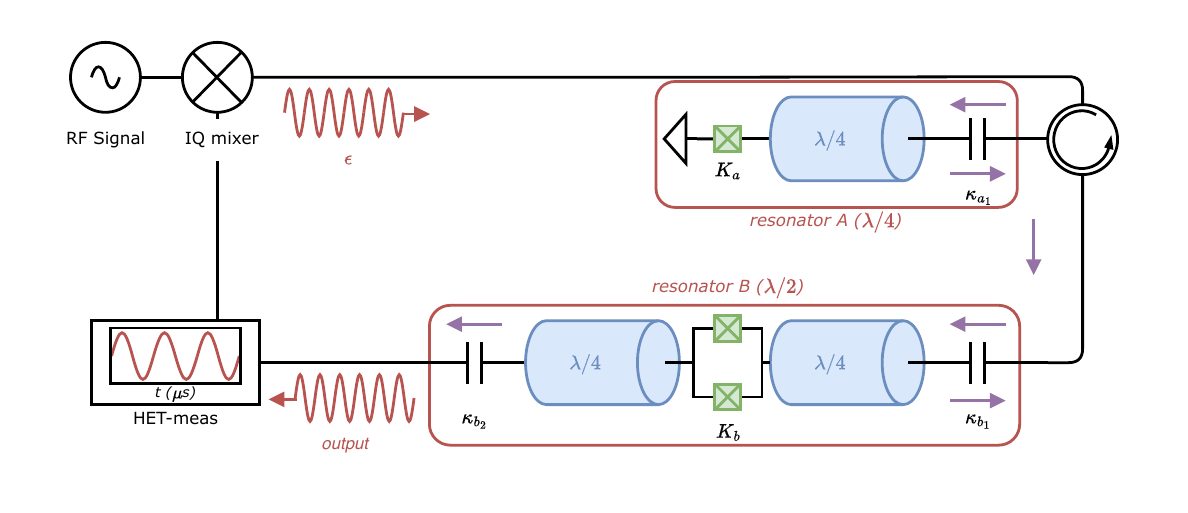}
\caption{An equivalent measurement based feedback scheme. The coherent feedback channel is replaced by a heterodyne measurement producing two currents that are used for IQ-modulation of the coherent field driving the $\lambda/4$ resonator.}
\label{meas-feedback1}
\end{figure}

In this section, we derive the equivalent measurement-based feedback model and compare it to the one for coherent feedback. The scheme is shown in shown in Fig.~\ref{meas-feedback1}. Note that operator hats are omitted for the remainder of this section.
The field exiting to the left of cavity B is subject to heterodyne detection. This produces two quadrature stochastic currents $J_x(t), J_y(t)$ that satisfy the stochastic differential equations
\begin{align}
    J_x(t) & = \kappa_{b,o}\langle b+b^\dagger\rangle_c+\sqrt{2\kappa_{b,o}} dW_x,\\
    J_y(t)  & =   -i\kappa_{b,o}\langle b-b^\dagger\rangle_c+\sqrt{2\kappa_{b,o}} dW_y.
\end{align}
A complex stochastic process $J_{het}(t) = \frac{1}{2}(J_x(t)+iJ_y(t))$ depends on the conditional quantum mean cavity field as 
\begin{equation}
    J_{het}(t)dt = \kappa_{b_o}\langle b\rangle_c(t)+dZ(t),
\end{equation}
where $dZ(t)=\sqrt{\kappa_{b,o}/2}(dW_x(t)+idW_y(t))$ is a complex valued Wiener process. The conditional mean is given by $\langle b\rangle_c(t) ={\rm Tr}[\rho_c(t) b] $ and the conditional state obeys the stochastic master equation \cite{wiseman_quantum_2009}
\begin{align}
    d\rho_c(t) & =  -i[H,\rho]dt +\kappa_a{\cal D}[a]\rho_c dt+\kappa_b{\cal D}[b]dt \nonumber\\
    & +{\cal H}[dZ^*(t) b]\rho_c.
\end{align}
The heterodyne current modulates the field input from the left using a IQ modulator. The feedback Hamiltonian is 
\begin{equation}
    H_{fb}= \lambda (J_x(t) X+J_{y}(t) Y),
\end{equation}
where $X  =  (a e^{-i\phi}+a^\dagger e^{i\phi})$, $Y  =  -i(a e^{-i\phi}-a^\dagger e^{i\phi})$ and $\lambda$ is a dimensionless parameter that accounts for the feedback strength. 

If we then average over all stochastic currents, the measurement-based feedback master equation is given by~\cite{wiseman_all-optical_1994}
 \begin{align}
 \label{meas-fb-me}
     \frac{d\rho}{dt} = &-i[H, \rho] + \kappa_b{\cal D}[b]\rho +\kappa_a{\cal D}[a]\rho \nonumber\\
     &-i\lambda\sqrt{\kappa_{b_{2}}\kappa_{a_{1}}}[ X,b\rho+\rho b^\dagger]-i\lambda\sqrt{\kappa_{b_{2}}\kappa_{a_{1}}}[ Y,-ib\rho+i\rho b^\dagger] \nonumber\\
     &+\lambda^2\kappa_{a_{1}}\kappa_{b_{2}}\left (2 {\cal D}[Y]\rho+2{\cal D}[X]\rho\right ) \nonumber\\
     &-\sqrt{\kappa_{a_{1}}\kappa_{b_{1}}}\left ([b^\dagger, a\rho]+[\rho a^\dagger,b]\right ).
 \end{align}
 Here, the Hamiltonian is 
    \begin{align}
            H  =& {\epsilon}(a+a^\dagger)+ \Delta_a a^\dagger a+\Delta_b b^\dagger b \nonumber\\
            &+  K_a a^{\dagger 2} a^{2}+K_b b^{\dagger 2} b^{2},
        \end{align} 
where we use the interaction picture. The semiclassical equations of motion are
\begin{align}
    \dot\alpha  =&  -\big( i \Delta_a + \frac{\kappa_a}{2} \big) \alpha  - 2 i K_a |\alpha |^2 \alpha-i{\epsilon} \nonumber \\
      &-2i\lambda\sqrt{\kappa_{a_{1}}\kappa_{b_{2}}}\ \beta e^{i\phi},\\
    \dot\beta  =&  - \left( i\Delta_b  + \frac{\kappa_b}{2} \right)  \beta - 2 i K_b |\beta|^2 \beta-\sqrt{\kappa_{a_{1}}\kappa_{b_{1}}} \alpha.
\end{align}
Apart from the driving term, these are the same as Eqs.~(\ref{eq:classical_a},\ref{eq:classical_b})
derived for the coherent feedback, if we choose $\phi=\pi/2, \ \lambda=1/2$ and neglect losses and relative phases. The resulting dynamics is the same and limit cycles occur for the same parameter values. 

\section{Noise analysis}
\label{sec: noise analysis}

We can convert the master equation to an equivalent stochastic process in an extended phase-space using the generalized P representation of Drummond and Gardiner \cite{WM}. The state of the two-mode system may be written as 
\begin{equation}
\label{gen-P}
\rho=\int_D \Lambda({\bm \alpha},{\bm \beta})P({\bm \alpha},{\bm \beta}) d\mu({\bm \alpha},{\bm \beta})
\end{equation}
where ${\bm \alpha}=(\alpha_a,\alpha_b),{\bm \beta}=(\beta_a,\beta_b)$. 
The non-diagonal coherent state projection operator is defined as 
\begin{equation}
\Lambda({\bm \alpha},{\bm \beta}) =\frac{|{\bm \alpha}\rangle\langle {\bm \beta}^*|}{\langle {\bm \beta}^*|{\bm \alpha}\rangle},
\end{equation} 
 and $d\mu({\bm \alpha},{\bm \beta})$ is an integration measure defining different representations.  The  Glauber-Sudarshan P function assumes that the stochastic dynamics never leaves the classical subspace ${\bm \beta}={\bm \alpha}^*$, and this is not the case when a Kerr nonlinearity is included. 

\subsection{Coherent feedback}
The Fokker-Planck equation is
 \begin{align}
\frac{\partial P}{\partial t} = & \sum_{\nu=a,b}\left[ \frac{\partial}{\partial \alpha_\nu} (\kappa_\nu' \alpha_\nu + 2i K_\nu \alpha_\nu^2 \beta_\nu) \right . \\\nonumber
&+  \frac{\partial}{\partial \beta_\nu} (\kappa_\nu'^* \beta_a - 2i K_\nu \beta^2 \alpha_\nu)\\\nonumber
& - i K_\nu \frac{\partial^2}{\partial \alpha_\nu^2}\alpha_\nu^2 + i K_\nu \frac{\partial^2}{\partial \beta_\nu^2} \beta_\nu^2 \\\nonumber 
& +\bar{\epsilon}\frac{\partial}{\partial \alpha_b}+\bar{\epsilon}^*\frac{\partial}{\partial \beta_b}\\\nonumber
&  \left .  +g_b(\alpha_a\frac{\partial}{\partial \alpha_b}+\beta_a\frac{\partial}{\partial \beta_b})+g_a(\alpha_b\frac{\partial}{\partial \alpha_a}+\beta_b\frac{\partial}{\partial \beta_a}) \right ]P,
\end{align}
where $\kappa_a' =-i\Delta_a+\kappa_a/2$ and $\kappa_b' =-i\Delta'_b+\kappa_b/2$. Note that the coherent feedback term does not add any second order derivatives. This indicates that it does not add any additional noise. The noise is determined by the quantum noise arising from the Kerr nonlinearities alone.

\subsection{Measurement-based feedback}
The terms in Eq.~(\ref{meas-fb-me}) that contribute classical diffusion noise due to measurement-based feedback are   
\begin{align}
   \frac{d\rho}{dt}  = &\ldots + \lambda^2\kappa_{a_{1}}\kappa_{b_{2}}\left (2 {\cal D}[Y]\rho+2{\cal D}[X]\rho\right ) \nonumber \\
    = &\ldots -\lambda^2\kappa_{a_{1}}\kappa_{b_{2}}([X,[X,\rho]]+[Y,[Y,\rho]])
\end{align}
In this form, we see that these terms drive a classical diffusion process in $X$ and $Y$ at the rate $\Gamma=\lambda^2\kappa_{a_{1}}\kappa_{b_{2}}$. This is absent from the coherent feedback master equation. Such terms contribute second order derivatives to the Fokker-Planck equation of the form
\begin{align}
    \frac{\partial P}{\partial t}  =\ldots &  +\Gamma \left (\frac{\partial}{\partial \alpha_a}+\frac{\partial}{\partial \beta_a}\right )^2P \nonumber \\
    &+\Gamma \left (\frac{\partial}{\partial \alpha_a}-\frac{\partial}{\partial \beta_a}\right )^2P 
\end{align}
We conclude that the coherent feedback process must necessarily be less noisy than the measurement-based feedback process given otherwise equivalent system parameters.

\section{Experimental Realisation}
\label{sec: experimental realisation}

\subsection{Nonlinear microwave resonators}
To show the feasibility of coherent feedback and its application for a quantum clock, we experimentally realise the feedback circuit using a system of coupled nonlinear superconducting resonators. While the proposed clock can in principle be implemented in other platforms such as photonic or optomechanical systems, our approach offers multiple advantages including control over certain design parameters, in particular, the coupling rates and the Kerr nonlinearities. The geometric coupling rates are determined by the dimensions of the capacitors connecting the resonators to the transmission lines and can be modelled using a finite-element simulation software.
To implement the Kerr nonlinearities, we terminate the $\lambda/4$-resonator by a single Josephson junction shorted to ground and intersect the $\lambda/2$-resonator B by a SQUID, which enables us to tune the resonance frequency $\omega_b$ in situ and, consequently, the detuning between the resonators. The necessary magnetic flux is supplied by a current-threaded copper coil attached to the bottom of the sample holder. Although this nonlinearity is not strictly necessary for the emergence of limit cycles, it adds an additional degree of freedom to the system to reach the necessary parameter regime. This is especially important since we do not have direct control over the losses and relative phases between the resonators, which affect the range in which limit cycles emerge. Additionally, the simulated model is semiclassical and does not include noise, so deviations from the predicted regime are expected. In the context of a clock, the nonlinearity in the oscillator system is equivalent to having the oscillator of a pendulum clock driven beyond its linear approximation regime.

The required nonlinearities are small compared to the resonator loss rates $K/\kappa \ll 1$, similar to the regime used for JPAs and tunable couplers. This enables us to discard the cross-Kerr and beamsplitter-like interaction terms between different modes and leaves us with the self-Kerr terms in Eq.~(\ref{eq:clock_hamiltonian}).
% As the bistable regime is highly sensitive to the specific parameters, semiclassical simulations were performed solving the equations of motion \ref{eq:classical_xa}-\ref{eq:classical_yb} to get estimates for the required system parameters to enable the existence of limit cycles in the megahertz regime.

Since we need to design our system parameters as accurately as possible in order to reach the regime where limit cycles occur, it is necessary to find explicit expressions for the resonator coupling rates, the Kerr nonlinearities and the resonator frequencies.
Following the approach detailed in Ref.~\cite{bourassa_josephson_2012}, we start from the Lagrangian for a coplanar waveguide (CPW) of length $d$ intersected or terminated by a Josephson junction and find the boundary conditions at the open and closed ends of the CPW and at the junctions.
Inserting the standard ansatz for travelling modes into the boundary conditions at the junction results in a transcendental equation for the wave numbers. Using that the junction capacitances $C_{J}$ are much smaller than the capacitance of the coplanar waveguides $C_J/(C_0 d_{a,b}) \ll1$, where $C_0$ is the capacitance of the waveguide per unit length and $d_{a,b}$ are the geometric lengths of the resonators, results in the following approximate expressions for the fundamental modes
\begin{equation}
    k^{(0)}_a l_a \approx \frac{\pi}{1+\gamma_a}, \qquad k^{(0)}_b l_b \approx \frac{\pi}{2\left(1+\gamma_b\right)}.
\end{equation}
Here, we define the inductive participation ratio as 
\begin{equation}
\gamma_{i} = \frac{L_{\text{Ji}}}{L_0 d_{i}} \ll 1\label{eq: gamma inductive},
\end{equation} 
where $L_{\text{Ji}}$ is the Josephson inductance and $L_{0}$ the geometric inductance per unit length of resonators A ($i=a$) and B ($i=b$).
This allows us to write analytical expressions for the effective resonator frequencies (dressed by the junction inductances) as
\begin{equation}
   \omega_a \approx \frac{\omega^{(0)}_a}{1+\gamma_a}, \qquad \omega_b(F) \approx \frac{\omega^{(0)}_b}{1+\cfrac{\gamma_b}{|\cos{(F)}|}}
    \label{eq: eff freq},
\end{equation}
where $\omega^{(0)}_{i}/2\pi = (L_{0} C_0)^{-1/2}$, $F=\pi \Phi/\Phi_0$ with $\Phi$ the external magnetic flux threading the SQUID loop  and $\Phi_0$ is the flux quantum. While both resonator frequencies are reduced by the presence of the junctions, applying an external flux to resonator B allows us to further decrease the frequency. Choosing $\omega_{b}(0)>\omega_{a}$ enables us to tune the resonators to the desired detuning and compensate for slight frequency deviations between the designed and the fabricated circuit. The geometric coupling rates are modified by the junctions as well, resulting in the following expressions for the effective coupling rates 
\begin{equation}
\kappa_{a_{1}}\approx \frac{\kappa^{(0)}_{a_{1}}}{1+4\gamma_{a}},
\qquad
\kappa_{b_{1,2}}(F) \approx \frac{\kappa^{(0)}_{b_{1,2}}}{1+4\cfrac{\gamma_b}{|\cos{(F)}|}},
\label{eq: kappa_a and kappa_b}
\end{equation}
where $\kappa^{(0)}_{i_{1,2}}$ ($i=a,b$) are the geometric coupling rates of the resonators. These differences between the bare and effective resonator frequencies and coupling rates have to be taken into account when designing the circuit.

We obtain the Hamiltonians for the nonlinear resonators by performing a Legendre transformation and substituting the creation and annihilation operators introduced in Eq.~(\ref{eq:clock_hamiltonian}). Expanding the nonlinear Josephson junction potential to second order, an approximation valid for small nonlinearities, and changing to the frame of the drive at frequency $\omega_\epsilon$, yields 
\begin{align}
\hat{H}_a &= \Delta_a \hat{a}^\dagger \hat{a} + K_{a} \hat{a}^{\dagger2} \hat{a}^2, \\
\hat{H}_b(F) &= \Delta_b(F) \hat{b}^\dagger \hat{b} + K_{b}(F) \hat{b}^{\dagger2} \hat{b}^2
\label{eq:hamiltonian res}
\end{align}
with the detunings $\Delta_{a,b}$ as defined in Eq.~(\ref{eq: cav detuning}) and the flux dependencies explicitly stated. The coefficients of the fourth-order terms in $H_{a,b}$ are the self-Kerr nonlinearities and given by
\begin{equation}
    K_a = -\frac{E_{Ja}}{4 \hslash} \left(\frac{\varphi_{zpf}}{\varphi_0} \right)^4 \cos \left(\frac{\pi }{2 (1+\gamma_a)}\right),
    \label{eq: kerr coeff a}
\end{equation}
\begin{equation}
    K_b(F) = -\frac{E_{Jb}}{2 \hslash} \left(\frac{\varphi_{zpf}}{\varphi_0} \right)^4   \cos \left(\frac{\pi }{2 \left(1+ \cfrac{\gamma_b}{|\cos{(F)}|} \right)}\right),
    \label{eq: kerr coeff b}
\end{equation}
where $\varphi_{\text{zpf}} = \sqrt{\hslash/(2 \omega_{a,b}^{(0)} C_{0}d_{a,b})}$ and $\varphi_0 = \Phi_0/(2\pi)$ are the flux zero-point fluctuation and the reduced flux quantum, respectively, and $E_{Ji} = \varphi_0^2/L_{Ji}$ with $i=(a,b)$. We solve Eq.~(\ref{eq: kerr coeff a}) and Eq.~(\ref{eq: kerr coeff b}) numerically to find the Josephson energies corresponding to the required Kerr nonlinearities. Having explicit expressions for all variables in Eqs.~(\ref{eq:classical_a}, \ref{eq:classical_b}), we are able to choose suitable fabrication parameters for the regime where stable limit cycles appear in the semiclassical simulations.

The device is fabricated using a standard process for superconducting circuits: First, the resonators are patterned in Aluminium on a Silicon substrate using electron-beam lithography and wet etching. Next, the Josephson junctions for the single junction and the SQUID are deposited via double-angle evaporation of aluminium and dynamic oxidation of aluminium oxide. Finally, the device is packaged and mounted in a dilution refrigerator and cooled to a temperature of $\qty{70}{\milli\kelvin}$. Details about the measurement setup and the data acquisition are given in Appendix \ref{app:data_acquisition}.

%The mechanical properties for the coplanar waveguides, i.e. the resonance frequencies, the geometric capacitances and inductances are determined analytically using the standard expressions (e.g. given in \cite{pozar2021microwave}). The resonators are coupled to the transmission line via interdigital capacitors based on DC simulations in HFSS Maxwell.
\subsection{Characterisation}

\begin{figure}
    \begin{subfigure}[b]{\linewidth}
        \includegraphics[width=\linewidth]{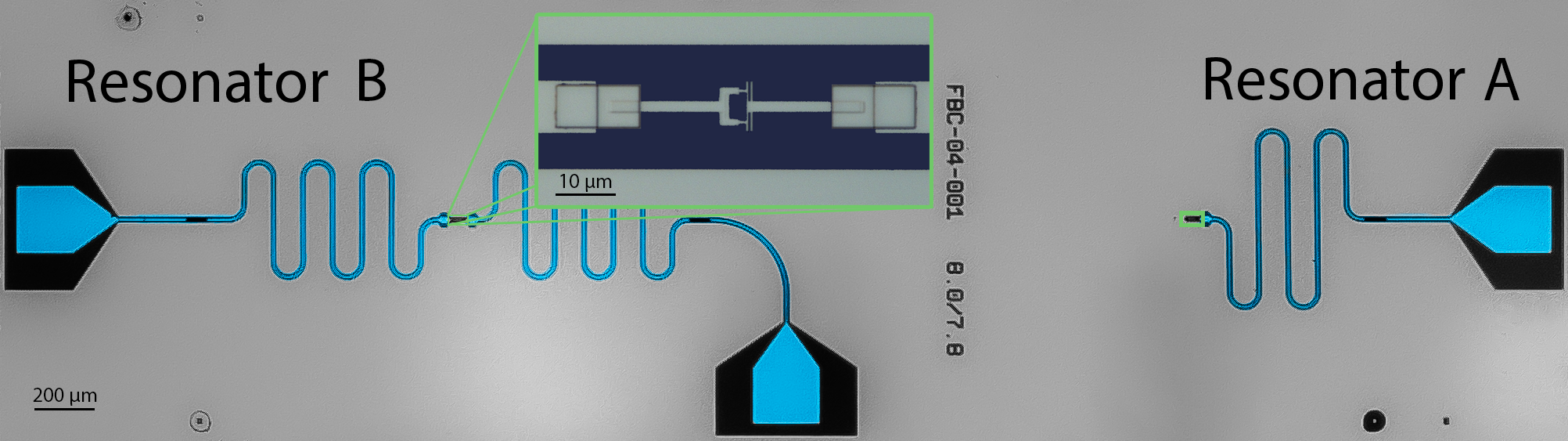}
        % \caption{\label{fig: fb loop circuit}}
        
    \end{subfigure}
    \caption{\label{fig:circuit schematic} Optical micrograph of the feedback circuit. The nonlinear resonators A (B) are shorted to ground (intersected) by a single junction (SQUID) with the location indicated by the green rectangle. A close-up of the SQUID, fabricated using double-angle evaporation is shown in the inset. The resonators are capacitively coupled to the the drive and readout circuitry via superconducting transmission lines (see the schematic in Fig.~\ref{fig: fb loop schematic}).}
\end{figure}
\begin{figure}
    \begin{subfigure}[b]{\linewidth}
        \includegraphics[width=\linewidth]{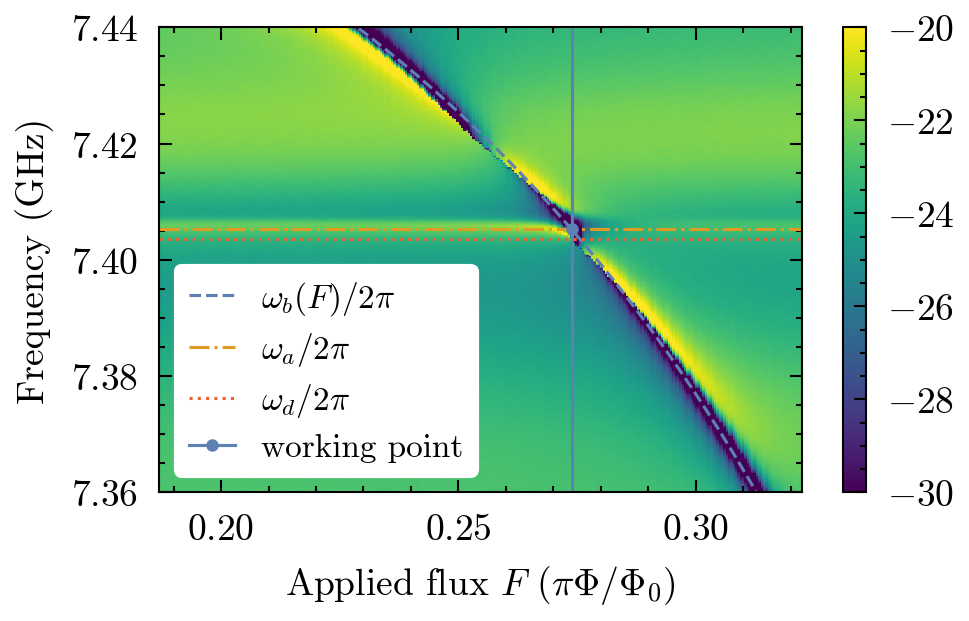}
        % \caption{\label{fig: resonance fit}}
    \end{subfigure}
    \caption{\label{fig: resonator characterisation} Partial flux sweep for resonator B. The resonance frequency $\omega_b$ decreases as the applied flux, increases according to Eq.~(\ref{eq: eff freq}). The fixed frequency of resonator a $\omega_a$ and the set drive frequency $\omega_\epsilon$ are shown in orange (dash-dotted) and green (dotted), respectively, and the flux working point at $F=0.274\pi$, used for the limit cycle measurements in this paper, is indicated by the red vertical line.}
\end{figure}
We first characterise both CPW resonators in a separate cool down without the feedback loop by applying an attenuated microwave signal directly to the input port of each resonator and measuring the response in reflection. From fitting the signal to an inverse Lorentzian, we extract the internal and external quality factors that give the loss rates $\kappa_{a,b}$ of the feedback clock system. 
To determine the Kerr nonlinearities $K_{a,b}$, we extract the inductive participation ratios $\gamma_{a, b}$ which are defined in Eq.~(\ref{eq: gamma inductive}).
While $L_0 d_{a,b}$  is determined by the dimensions of the coplanar waveguide resonator which is controlled by the resonator geometry, the Josephson inductances $L_{Ja,b}$ are determined by the junction parameters, i.e. the junction area and the oxide thickness. Due to uncertainties in the fabrication process, these values can deviate from the design.

An initial estimation for $L_{J,a}$ is obtained from measuring the critical current at room temperature using the Ambegaokar-Baratoff relation. At cryogenic temperatures, a more precise way to determine $L_{Ja}$ is to measure the effective frequency $\omega_a$ and get $\gamma_a$ from Eq.~(\ref{eq: eff freq}), using the design values for $L_a$, $l_a$ and $\omega_a^{0}$. 
For the tunable resonator B, we sweep the external magnetic flux resulting in a $\pi$~-periodic modulation of the resonance frequency $\omega_b(F)$, as partially shown in Fig.~\ref{fig: resonator characterisation}.
Fitting such a curve to Eq.~(\ref{eq: eff freq}) and using the flux periodicity yields the SQUID inductance $L_{J,b}$.
From $\gamma_{a,b}$, we finally calculate $K_a$ and $K_b(F)$ using Eqs.~(\ref{eq: kerr coeff a}) and (\ref{eq: kerr coeff b}), respectively.

\begin{table}[b]
\begin{ruledtabular}
\begin{tabular}{lrl}
Parameter & Value & Unit \\ \hline \hline
$F=\pi\Phi/\Phi_0$ & $0.274\pi$ & $-$  \\ \hline 
$\omega_a/2 \pi$ & 7.4060 & GHz  \\ \hline 
$\omega_b(0)/2 \pi$ & 7.4960 & GHz   \\ \hline 
$\omega_b(F)/2 \pi$ & 7.4053 & GHz  \\ \hline 
$\omega_\epsilon/2 \pi$ & 7.4035 & GHz   \\ \hline 
$\Delta_a/2 \pi$ & 1.80 & MHz   \\ \hline 
$\Delta_b(F)/2 \pi$ & 1.80 & MHz   \\ \hline 
$K_a/2 \pi$ & -0.01 & MHz \\ \hline 
$K_b(0)/2 \pi$ & -0.01 & MHz  \\ \hline 
$K_b(F)/2 \pi$ & -0.03 & MHz   \\ \hline 
$\kappa_a/2 \pi$ & 3.20 & MHz   \\ \hline 
$\kappa_{a_{\mathrm{int}}}/2 \pi$ & 0.11 & MHz   \\ \hline 
$\kappa_{b1,2}(0)/2 \pi$ & 2.64 & MHz  \\ \hline 
$\kappa_{b1,2}(F)/2 \pi$ & 2.52 & MHz  \\ \hline 
$\kappa_{b_{\mathrm{int}}}/2 \pi$ & 1.64 & MHz   %\\ \hline 
% $\varphi$ & 0 & rad     \\ \hline 
% $\eta$ & $\sqrt{0.99}$ &  $-$    \\ \hline 
% $\epsilon^2$& $2\times 10^{10}-2 \times 10^{11}$ & photons/s
\end{tabular}
\end{ruledtabular}
\caption{\label{tab: experimental_parameters} System parameters of the superconducting circuit used for the measurements. The values are determined from the characterisation of the resonators. See the main text for details.}
\end{table}

After the resonators are characterised, we add the remaining parts of the feedback circuit as shown in Fig.~\ref{fig:circuit schematic}. Driving the system through the directional coupler and measuring the signal reflected off resonator B and transmitted via the circulator shows both resonators appear as dips in the transmission amplitude (dashed blue line and dash-dotted orange line in Fig.~\ref{fig: resonator characterisation}). From this, we identify the flux value where the resonance frequencies coincide. We expect limit cycles to appear near this working point.

The drive power applied to the circuit is determined by the total attenuation in the microwave input line and related to the photon number rate $\epsilon^2$ by
$P_{\text{dBm}} = 10\log_{10}{(\hslash \omega_\epsilon \epsilon^2)}+30$, and set to the range where the limit cycles are expected to appear. The relative phase between the resonators, $\varphi_{1,2}$, and the loss coefficients, $\eta_{1,2}$, cannot be measured directly and are kept as free parameters that can be fixed by comparing the simulations to the experimental results.

\section{Results}
\label{sec: results}

After sweeping the power and frequency of the drive and the frequency of resonator B around resonator A within the range suggested by the semiclassical equations of motion, we observe the emergence of limit cycles for a range of parameters. For the following measurements, we choose settings that allow for a clear observation of the limit cycles through a wide range of driving powers. The system parameters used are shown in Table~\ref{tab: experimental_parameters}.

The signal emitted from the clock system through the bottom circulator in Fig.~\ref{fig: fb loop schematic} comprises primarily of the photons reflected from resonator B and needs to be amplified to be registered at room temperature. This is achieved through an amplification chain containing a quantum-limited amplifier, a high-electron mobility transistor at 4 Kelvin and two room-temperature amplifiers. The signal is then processed using a heterodyne detection scheme consisting of analogue IQ-demodulation mixing the signal with a local oscillator shifted by $\qty{25}{\mega\hertz}$ from the drive frequency and subsequent digital filtering and downconversion to DC using an analogue to digital converter. We end up with a time series of the quadrature components of the signal, which we can analyse directly or process further.

To explore the properties of the limit cycles, we move to the frequency domain, which, in contrast to the time domain, allows taking averages to increase the signal-to-noise ratio without suppressing the signal itself. We do this by taking the absolute square of the Fourier transform which gives the energy spectral density (ESD). Here, the limit cycles appear as sidebands centred around the driving frequency at $\qty{0}{\hertz}$.

\begin{figure}
    \begin{subfigure}[b]{\linewidth}
        \includegraphics[width=\linewidth]{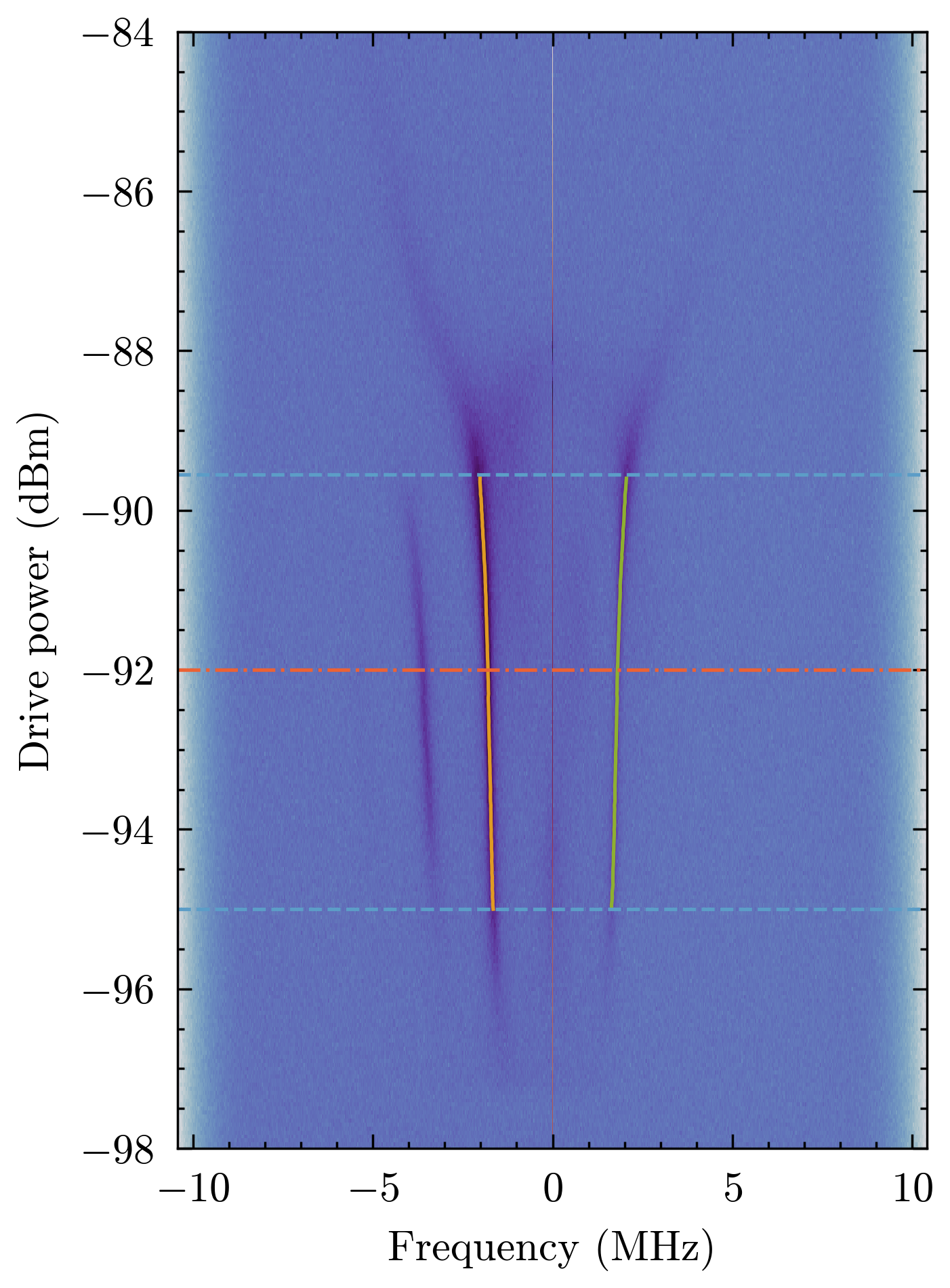}
        \label{fig: cd3-esd_noiseless-lines}
    \end{subfigure}
    
    \begin{subfigure}[b]{\linewidth}
        \includegraphics[width=\linewidth]{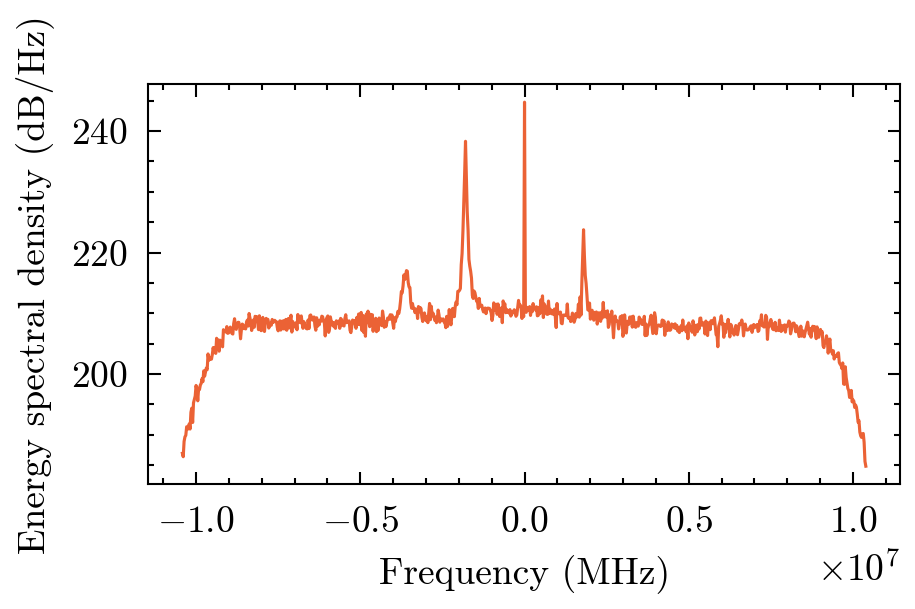}
        \label{fig: cd3-esd_noiseless-cutout}
    \end{subfigure}
\caption{\label{fig: cd3-esd_noiseless} \textbf{(Top)} ESD of a drive power sweep that results in the formation of limit cycles, visible as sidebands from $\sim \qty{1.5}{\mega\hertz}$. The dashed blue lines indicate the lower and upper boundaries for the drive power used in Fig.~\ref{fig: sidebands}.  The drive signal is visible as a thin line at $\qty{0}{\hertz}$. \textbf{(Bottom)} Cutout of the above sweep at $-92$~dBm (red dash-dotted line). Visible are the drive in the centre, the sidebands at $\sim \qty{2}{\mega\hertz}$, asymmetric in amplitude and linewidth, the second harmonic of the negative sideband, and the digital low-pass filter at $\qty{10}{\mega\hertz}$.
}
\end{figure}

For the system parameters in Table~\ref{tab: experimental_parameters}, a drive power sweep averaged from $100$ repetitions is depicted in the top of Fig.~\ref{fig: cd3-esd_noiseless}. Bifurcation occurs at around $-96$~dBm and limit cycles emerge at a frequency of around $\qty{1.5}{\mega\hertz}$ ($\qty{0.6}{\micro\second}$). Increasing the drive power increases the limit cycle frequency up to around $\qty{2.0}{\mega\hertz}$ ($\qty{0.5}{\micro\second}$) at $-89$~dBm, beyond which bifurcation doubling splits the limit cycles in two branches.
The bottom of Fig.~\ref{fig: cd3-esd_noiseless} depicts a cutout of the ESD at $-92$~dBm and shows several distinct features of the sidebands: the significant asymmetries in amplitude and linewidth between the positive and negative sideband and the second harmonic visible in the negative sideband starting from around $\qty{-3}{\mega\hertz}$. Neither of these features are captured by our semiclassical model, and while the second harmonic is likely due to higher-order nonlinearities becoming relevant, the origin of the asymmetry in the sidebands has not yet been identified.

\begin{figure}
\centering
    \includegraphics[width=\linewidth]{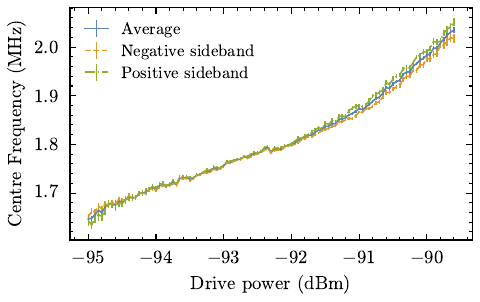}    \includegraphics[width=\linewidth]{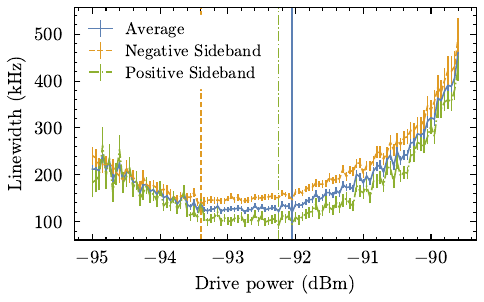}
    % \includegraphics[width=\textwidth]{Figures/cd3-noisy_linewidths.png}
    % \caption{LCs driven by a (white) noisy signal. While both the drive signal and the LCs are broadened in frequency, the LCs appear to be more robust to the noise.}
    % \label{fig: cd3-noisy_linewidths}
\caption{\label{fig: sidebands}Sideband frequencies (\textbf{top}) and linewidths (\textbf{bottom}) in the driving power range indicated in Fig.~\ref{fig: cd3-esd_noiseless}. The centre frequency and linewidths are obtained by fitting to a Lorentzian. The error bars correspond to the standard deviations of the fit parameters.}
\end{figure}

To show the behaviour of the limit cycles with increasing drive power, we extract the centre frequency and the linewidths of the sidebands by fitting the peaks to a Lorentzian. The change of the frequencies and the linewidths are shown in Fig.~\ref{fig: sidebands}. As expected, the frequency increases with driving power while, in the low-power regime, the linewidth decreases with power as predicted by \cite{erker_autonomous_2017}. The following increase in the linewidth can be likely attributed to higher-order nonlinearities indirectly manifested by the appearance of second harmonics as discussed above.

\subsection{\label{sec:quantum clock}Application as a quantum clock} 
\begin{figure}
    \begin{subfigure}[b]{\linewidth}
        \includegraphics[width=\linewidth]{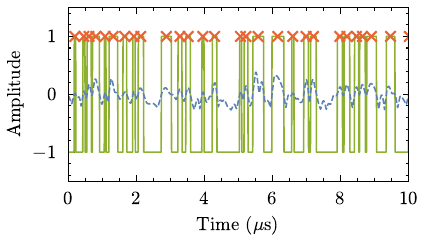} 
        \caption{\label{fig: clock tick distribution}}
    \end{subfigure}
    \begin{subfigure}[b]{\linewidth}
        \includegraphics[width=\linewidth]{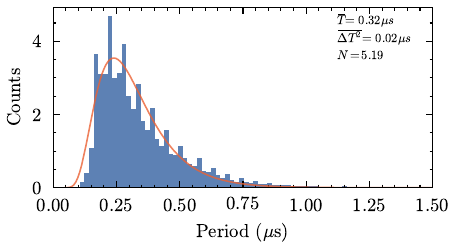}
        \caption{\label{fig:cd3-time_traces-wald}}
    \end{subfigure}
\caption{(\textbf{Top}) Cutout of a time series taken for $\qty{12.8}{\milli\second}$ at $-93.5$~dBm drive power and with a digital low-pass filter at $\qty{4}{\mega\hertz}$ to remove high-frequency noise (dotted blue line). To extract clock ticks, we zero-centre the data and apply a sign function (green line). The period of the clock ticks is then given by the difference between the peaks (orange crosses). 
(\textbf{Bottom}) Extracted clock ticks fitted to an inverse Gaussian (Wald) distribution with the mean $\overline{T}$, variance $\overline{\Delta T^2}$ and corresponding accuracy. The $R^2$ factor of the fit is 0.93.}
\end{figure}

A clock directly benefits from the reduced noise generated by coherent feedback compared to a measurement-based feedback approach, as the linewidth of the clock signal in frequency is directly related to the accuracy of the timing signal in the time domain. 
To generate a timing signal, we continuously measure the heterodyne output current of the feedback circuit which gives a complex signal in the time domain $s(t)=I(t)+i Q(t)$ with the in phase and quadrature currents $I(t)$ and $Q(t)$, respectively. The clock ticks are generated from the amplitude 
\begin{equation}
A(t)=\sqrt{|s(t)|^2}    
\end{equation} by normalising and zero-meaning it and applying the sign function (Fig.~\ref{fig: clock tick distribution}). 

The differences in time between the peaks yield the clock tick period distribution which is fitted to an inverse Gaussian (Wald) distribution 
\begin{equation}
    W(T,\alpha,\lambda) = \sqrt{\frac{\lambda}{2\pi}} T^{-3/2} \exp \left [-\frac{\lambda}{2\alpha^2 T} (T-\alpha)^2\right ], \quad t\geq 0,
\end{equation}
where $\alpha,\lambda$ are positive real parameters ($\lambda$ is called the spread parameter). From the mean $\overline{T}  = \alpha$ and the variance $\overline{\Delta T^2}  = \alpha^3/\lambda$, we get the clock accuracy $N$ via
\begin{equation}
    N = \frac{\overline{T} ^2}{\overline{\Delta T^2} } = \frac{\lambda}{\alpha}.
    \label{eq:accuracy}
\end{equation}
For time traces taken at detunings $\Delta_a = \qty{2.0}{\mega\hertz}$ and $\Delta_b = \qty{0.8}{\mega\hertz}$, this results in $N\approx 5.1$, as depicted in Fig.~\ref{fig:cd3-time_traces-wald}. While this is better than a random clock which, by definition, has $N=2$, it is many orders of magnitude lower than most practical clocks. However, the goal of our experiment is not to surpass existing clocks but to show the implementation of a new feedback scheme which may be used in practical devices to overcome the limitations of the clocks with classical feedback.

Since the linewidths of the sidebands are related to the clock tick distribution by 
\begin{equation}
\gamma = \frac{\sigma^2}{\mu^2},
\label{eq: clock sideband relation}
\end{equation} 
where $\mu$ is the amplification rate and $\sigma$ is the variability of noise (see \cite{Aminzare}), they can be used as an additional figure of merit. For our system settings, they reach minima of 136.34~kHz and 92.01~kHz, whereas the average has a minimum of 120.54~kHz, indicated by the vertical lines in the bottom of Fig.~\ref{fig: sidebands}. A derivation of Eq.~(\ref{eq: clock sideband relation}) is given in Appendix \ref{appendix1}.

\subsection{\label{sec:autonomous quantum clock} Towards an autonomous quantum clock} 
In the previous sections, we drove the system with a coherent drive which can be used as a clock by itself. To understand the effect of the coherence of the drive on the accuracy of our clock, we replace the coherent signal with a noisy driving field. To controllably reduce the coherence of the drive, we modulate the frequency of the drive signal using a white noise signal provided by a function generator with an upper cut-off frequency of $\qty{500}{\kilo\hertz}$. The effect of the frequency noise in the driving field results in the broadening of both the drive signal peak and sidebands in the ESD spectrum.

\begin{figure}
    \centering
    \includegraphics[width=\linewidth]{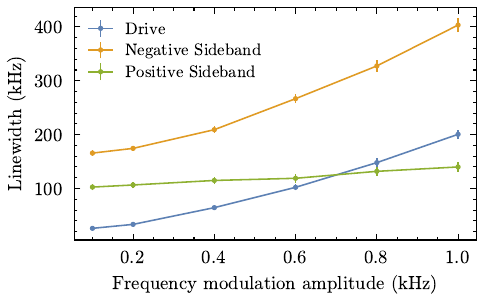}
    \caption{\label{fig:linewidths_noisy_drive}Linewidths of the noisy drive and sideband as a function of the frequency deviation of the drive signal. Above \qty{0.7}{\kilo\hertz}, the linewidth of the drive becomes wider than the left sideband, showing that coherence can be maintained even in the presence of noise. The error bars correspond to the standard deviation of the fit parameter.}
\end{figure}

However, we observe that for a certain drive power range, the linewidth of the positive sideband increases at a lower rate than the linewidth of the drive which leads to a crossover at the frequency modulation amplitude of $\qty{0.6}{\kilo\hertz}$. At stronger modulation the left sideband peak becomes narrower than the noisy drive peak, as shown in Fig.~\ref{fig:linewidths_noisy_drive}. This is consistent with the theory in Section \ref{sec: semiclassical dynamical analysis} which suggests that even a fully incoherent drive allows for limit cycles to form in the coherent feedback system.

Although the positive-frequency sideband remains narrow, the negative-frequency sideband broadens approximately at the same rate as the drive signal in the driving power range. To get an advantage in terms of clock performance, it is necessary to create the clock ticks using only the positive sideband signal, disregarding the negative-frequency sideband and the drive.
% We illustrate this process by switching back to the time domain:
% To extract the narrower sideband at $f_{LC}\approx \qty{1.87}{\mega\hertz}$ at the set drive power $ \qty{-93}{\dBm}$, $f_{LC}$ is shifted to zero via $s(t) \rightarrow s(t) e^{-i 2 \pi f_{LC} t}$ and a digital Butterworth low-pass filter of $10^{th}$ order from $ \qty{1}{\mega\hertz}$ is applied. From this, we extract a clock signal as in Section \ref{sec:quantum clock} where only the positive-frequency sideband signal contributes to the clock signal.

\section{Conclusion}
\label{sec: conclusion}

In this paper, we theoretically propose and experimentally realise a new type of quantum clock based on coherent feedback. Unlike traditional clock which rely on measurement-based feedback, our design leverages coherent feedback to suppress noise from quantum measurements, resulting in an improved clock accuracy. Employing semiclassical simulations, the system exhibits stable limit cycles which form the basis for periodic oscillations. We theoretically analyse the quantum noise characteristics of our system and show its inherent superior performance compared to measurement-based feedback. Experimentally, we realise this system using superconducting resonators with embedded Josephson junctions and microwave circulators. Our measurements confirm the emergence of limit cycles and show that the limit cycle oscillations are more coherent than the driving signal under noisy conditions, opening possibilities for implementation of autonomous quantum clocks. The clock’s accuracy is characterised in both frequency and time domains, showing promising results for future quantum metrology applications. While our clock accuracy is not directly comparable with state-of-the-art atomic clocks, we believe that coherent feedback can be used to reach higher accuracy and stability of clock in the future, thus, opening up a new avenue for quantum clock design. In addition to the application as a quantum clock, coherent feedback can be used in a wide range of quantum systems to improve performance and stability beyond the limit of measurement-based feedback.

\begin{acknowledgments}
L.A.M. acknowledges the support from the Brazilian Agency CAPES (Coordena{\c c}\~ao de Aperfei{\c c}oamento de Pessoal de N\'ivel Superior), Finance Code 88881.128437/2016-01. This project/research was supported by grant number FQXi-IAF19-04 from the Foundational Questions Institute Fund, a donor advised fund of Silicon Valley Community Foundation. We also acknowledge the support of the Australian Research Council Centre of Excellence for Engineered Quantum Systems (EQUS, CE170100009). The authors acknowledge the facilities, and the scientific and technical assistance, of the Australian Microscopy \& Microanalysis Research Facility at the Centre for Microscopy and Microanalysis, The University of Queensland. This work used the Queensland node of the NCRIS-enabled Australian National Fabrication Facility (ANFF).
\end{acknowledgments}

\appendix
\section{Limit cycles.}
\label{appendix1}
The central role of limit cycles in clock design is not widely appreciated. It is a non-trivial mathematical problem to determine the period of the limit cycle using centre manifold theory, done by parametrising the motion on the limit cycle by a phase variable, $\theta(t)$. In the absence of noise we would like this to be fixed by the parameters of the model. The  period is then defined as the time take for this variable to move through $2\pi$.  

However, noise causes phase diffusion and thus the time taken to move through $2\pi$ fluctuates from one cycle to the next.  The general theory (phase reduction) \cite{Aminzare} shows that the phase noise diffusion rate is slower the larger the limit cycle.  For example in the normal form of a limit cycle the equations of motion near a Hopf bifurcation are
\begin{align}
\dot{x}  = & [\mu x-\omega y-(x^2+y^2)y]dt \\
\dot{y}	 = & [\omega x+\mu y -(x+y^2)x]dt +\sigma dW(t),
\end{align}
where $\mu$ is the amplification  rate.  The phase variable on the limit cycle is defined by $x(t) =\sqrt{\mu }\cos(\theta(t))$, $p(t) =\sqrt{\mu }\sin(\theta(t))$. Phase unfolding then gives
\begin{equation}
\label{noisy-phase}
   d\theta(t) = \omega +\frac{\sigma}{\mu}dW(t), 
\end{equation}
i.e., the larger the limit cycle (larger $\mu$) the slower the phase noise.  Large limit cycles imply large heat dissipation as the work done in each cycle is dissipated as heat on each cycle, and the more dissipative, the better the clock \cite{erker_autonomous_2017}. A well know  example is the laser: the harder it is driven above  threshold, the smaller is the phase noise and the smaller the linewidth. 

\begin{figure}
\centering
\includegraphics[scale=0.5]{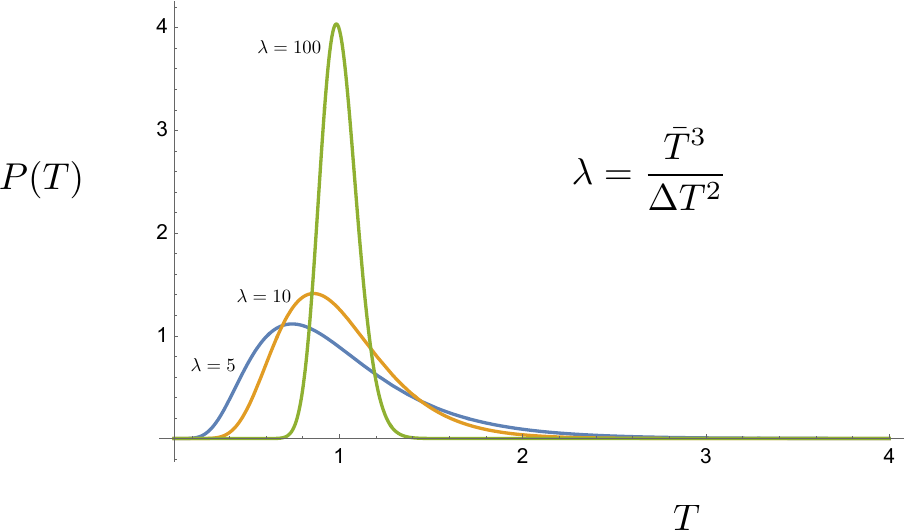}
\caption{Mean period $\bar{T}=1$ and three different values of the spread parameter, $\lambda$. Increasing $\lambda$ is decreasing noise. }
\label{Wald}
\end{figure}

We can find the statistics of the period as a first passage time problem: what is the probability for time $T$ taken for the phase to change by $2\pi$. This is given by the Wald or Inverse Gaussian distribution,
\begin{equation}
    W(T,\alpha,\lambda) = \sqrt{\frac{\lambda}{2\pi}} T^{-3/2} \exp \left [-\frac{\lambda}{2\alpha^2 T} (T-\alpha)^2\right ], \quad t\geq 0,
\end{equation}
where $\alpha,\lambda$ are positive real parameters ($\lambda$ is called the spread parameter). We give an example in Fig.~\ref{Wald}. The mean and variance are 
\begin{align}
    \overline{T}  = & \alpha\\
    \overline{\Delta T^2}  = & \frac{\alpha^3}{\lambda}.
\end{align}
The fractional uncertainty is given by  
\begin{equation}
\frac{\overline{\Delta T^2}}{\overline{T}^2}=\frac{\alpha}{\lambda}.
\end{equation}
For fixed $\alpha$ (fixed mean), this decreases as the spread parameter increases. 

In the case of a normal form limit cycle:
\begin{align}
\overline{T}  = & \frac{2\pi}{\omega}\\
\overline{\Delta T^2}  = & \frac{2\pi\sigma^2}{\omega^3\mu},
\end{align}	
fluctuations in period get smaller as the limit cycle gets bigger. 
If we define the frequency of the clock as
\begin{equation}
    f=\frac{1}{T},
\end{equation}
the moments for the frequency are determined by the Wald distribution as 
\begin{align}
   \overline{f}  = & \frac{1}{\alpha}+ \frac{1}{\lambda}\\
     \overline{\Delta f^2}   = & \frac{1}{\alpha\lambda}+\frac{2}{\lambda^2}.
\end{align}
The fractional uncertainty in the period is
\begin{equation}
    \frac{ \overline{\Delta f^2}}{\overline{f}^2}= \frac{\alpha\lambda+2\alpha^2}{(\alpha+\lambda)^2}\approx \frac{\alpha}{\lambda}
\end{equation}
for large $\lambda$ (low noise). Thus we have the typical result for a good clock
\begin{equation}
 \frac{ \overline{\Delta T^2}}{\overline{T}^2}= \frac{ \overline{\Delta f^2}}{\overline{f}^2}.
 \end{equation}

The power spectrum can be calculated directly from the SDE for the phase, Eq.~(\ref{noisy-phase}): Define the complex amplitude $z(t)$ such that 
\begin{equation}
 dz= iz(t)d\theta(t) =iz(t)(\omega dt+\frac{\sigma}{\mu}dW)
\end{equation}.
This is the noisy phase oscillator.  (See \cite{gardiner_handbook} page 104). The stationary two-time correlation function for heterodyne detection is given by 
\begin{equation}
    {\mathbb E}[z^*(0)z(\tau)]=e^{i\omega t-\frac{\sigma^2}{2\mu^2}|\tau|}.
\end{equation}
The noise power spectrum is the Fourier transform of this 
\begin{equation}
    P(\Omega) =\frac{1}{2\pi}\frac{\gamma}{\gamma^2+(\omega-\Omega)^2},
\end{equation}
where the line width is 
\begin{equation}
\gamma =\frac{\sigma^2}{\mu^2},
\end{equation}
a Lorentzian.

\section{Building the SLH triple for the clock}
\label{appendix2}

Here, we show how to obtain the Hamiltonian $\hat{H}$ and collapse operators $\hat{L}_i$ for the clock system using the SLH framework \cite{gough_quantum_2009, gough_series_2009}. Within this framework, we model each element as a triple of three terms: the scattering matrix $\mathbf{S}$, the collapse operators $\hat{L}_i$, and the Hamiltonian $\hat{H}$, so
\begin{equation}
    \mathbf{G}_j = \left(\mathbf{S}, \begin{bmatrix}
        \hat{L}_1 \\ \hat{L}_2 \\ \vdots \\ \hat{L}_n     \end{bmatrix}, \hat{H}  \right)
\end{equation}
Mathematically, $\mathbf{G_j}$ is the symbol that comprises S, $\hat{L}$, and $\hat{H}$ in a single of the $j$-component in the model.

\begin{figure}
    \includegraphics[width=\linewidth]{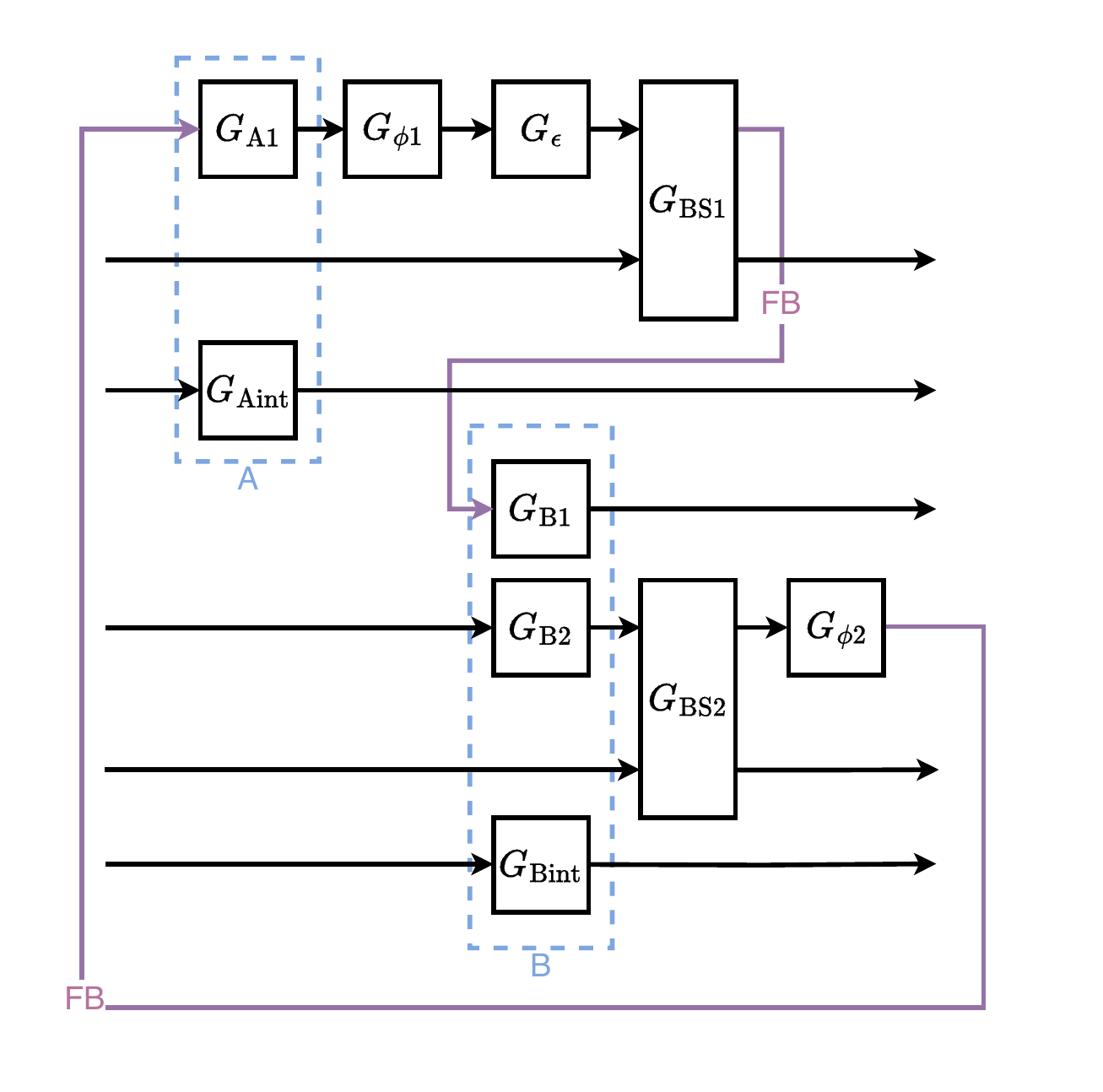}
    \caption{SLH circuit (with input and output states from left to right) used to derive the equations of motion for the superconducting clock. The internal losses are added as unguided modes while the losses between resonators are modelled by a beamsplitter with reflection coefficient $\eta$.}
    \label{fig:slh_scheme_alt}
\end{figure}

In terms of SLH building blocks, the superconducting clock can be decomposed as shown in Fig.~\ref{fig:slh_scheme_alt}. We note that the SLH decomposition is not unique \cite{combes_slh_2017} and there are multiple circuits that will describe the same physical system. 

First, we write the SLH triple for each of the component used in our decomposition:

The Hamiltonian for each resonator is given by:
\begin{equation}
    \hat{H}_i = K_{i} \hat{i}^{\dagger2} \hat{i}^2 + \Delta_{i} \hat{i}^{\dagger} \hat{i}.
\end{equation}
with the self-Kerr coefficients $K_{i}<0$, the detuning from the drive $\Delta_{i}$ defined in Eq.~(\ref{eq: cav detuning}) and $i=a,b$.

\begin{enumerate}
    \item Coherent Driving Field
    
    $ G_\epsilon = (\mathbb{I}_1, [\epsilon], 0) $,

    where $\mathbb{I}_1$ is the  $1\times1$ identity matrix and $\epsilon$ is the amplitude of the driving field.

    \item Resonator B - Side 1
    
    $G_{\mathrm{B}1} = (\mathbb{I}_1, [\sqrt{\kappa_{b_{1}}} \hat{b}], H_b)$.

    Although we the resonators can be modelled as single elements in the SLH framework, we decompose them into their constituents for our circuit model. 
%     The Hamiltonian of resonator B is given by:

%     \begin{equation}
%     H_b = K_{b} \hat{b}^{\dagger2} \hat{b}^2 + \Delta_{b} \hat{b}^{\dagger} \hat{b}
% \end{equation}

    \item Resonator B - Side 2

    $G_{\mathrm{B}2} = (\mathbb{I}_1, [\sqrt{\kappa_{b_{2}}} \hat{b}], 0)$
    
    \item Resonator A - Side 1

    $G_{\mathrm{A}1} = (\mathbb{I}_1, [\sqrt{\kappa_{a_{1}}} \hat{a}], H_a)$.

    The other side of resonator A is shorted to ground and not explicitly modelled.

    \item Resonators internal losses 
    
    \begin{align*}
    G_{{\mathrm{Aint}}} &= (\mathbb{I}_1, [\sqrt{\kappa_{a_{\mathrm{int}}}} \hat{a}], 0), \\ 
    G_{{\mathrm{Bint}}} &= (\mathbb{I}_1, [\sqrt{\kappa_{b_{\mathrm{int}}}} \hat{b}], 0)
    \end{align*}
    
    \item Beam Splitters

    $G_{\mathrm{BS}i} = \left(\mathbf{S}_{\mathrm{BS}i}, \begin{bmatrix}
        0 \\ 0
    \end{bmatrix}, 0\right)$,

    where i = 1,2 and 
    \begin{equation*}
        \mathbf{S}_{\mathrm{BS}_{i}} = \begin{bmatrix}
            \sqrt{1-\eta_i} & \eta_i \\ 
            -\eta_i & \sqrt{1-\eta_i}.
        \end{bmatrix}
    \end{equation*}
    We use two beam splitters to model losses due to the circulators between cavities A and B. 

    \item Phase shifter 

    $G_{\phi i} = \left([e^{(\phi_i)}], [0], 0\right)$,

    We used the phase shifter to model the phases acquired by the travelling field between resonators A and B and B and A.

    \item Padding

    $G_{\mathrm{pad}}=\left(\mathbb{I}_1, \left[0 \right], 0 \right)$, 
    
    Padding triplets are used to keep the dimensions of the concatenated elements consistent. Since they do not add any structure to the circuit, they are omitted in Fig.~\ref{fig:slh_scheme_alt}.
\end{enumerate}
Following the decomposition shown in Fig.~\ref{fig:slh_scheme_alt}, we write the following composition to describe our clock system:

\begin{align}
&\mathrm{FB}\{\mathrm{FB}\{[ (G_{\mathrm{BS}1} \boxplus G_{\mathrm{pad}}) \nonumber \\
&\triangleleft (G_\epsilon \boxplus G_{\mathrm{pad}} \boxplus G_{\mathrm{pad}}) \nonumber\\
&\triangleleft (G_{\phi 1} \boxplus G_{\mathrm{pad}} \boxplus G_{\mathrm{pad}}) \nonumber\\
&\triangleleft (G_{\mathrm{A}1} \boxplus G_{\mathrm{pad}} \boxplus G_{\mathrm{Aint}})] \\
&\boxplus [(G_{\mathrm{pad}} \boxplus G_{\mathrm{BS}2} \boxplus G_{\mathrm{pad}}) \nonumber\\
&\triangleleft (G_{\mathrm{pad}} \boxplus G_{\phi 2} \boxplus G_{\mathrm{pad}} \boxplus G_{\mathrm{pad}}) \nonumber\\
&\triangleleft (G_{\mathrm{B}1} \boxplus G_{\mathrm{B}2} \boxplus G_{\mathrm{pad}} \boxplus G_{\mathrm{Bint}})], 0, 3\}, 3, 0\} \nonumber
\end{align}
where the feedback (FB), series ($\triangleleft$) and concatenation ($\boxplus$) operations for SLH triples are defined as in Refs. \cite{gough_series_2009, gough_quantum_2009}. This is the decomposition used to construct the Hamiltonian (equation \eqref{eq:clock_hamiltonian}) and collapse operators (equations \eqref{eq:collapse_op}) for the superconducting clock system.

\section{Data acquisition}
\label{app:data_acquisition}
\begin{figure}
    \includegraphics[width=\linewidth]{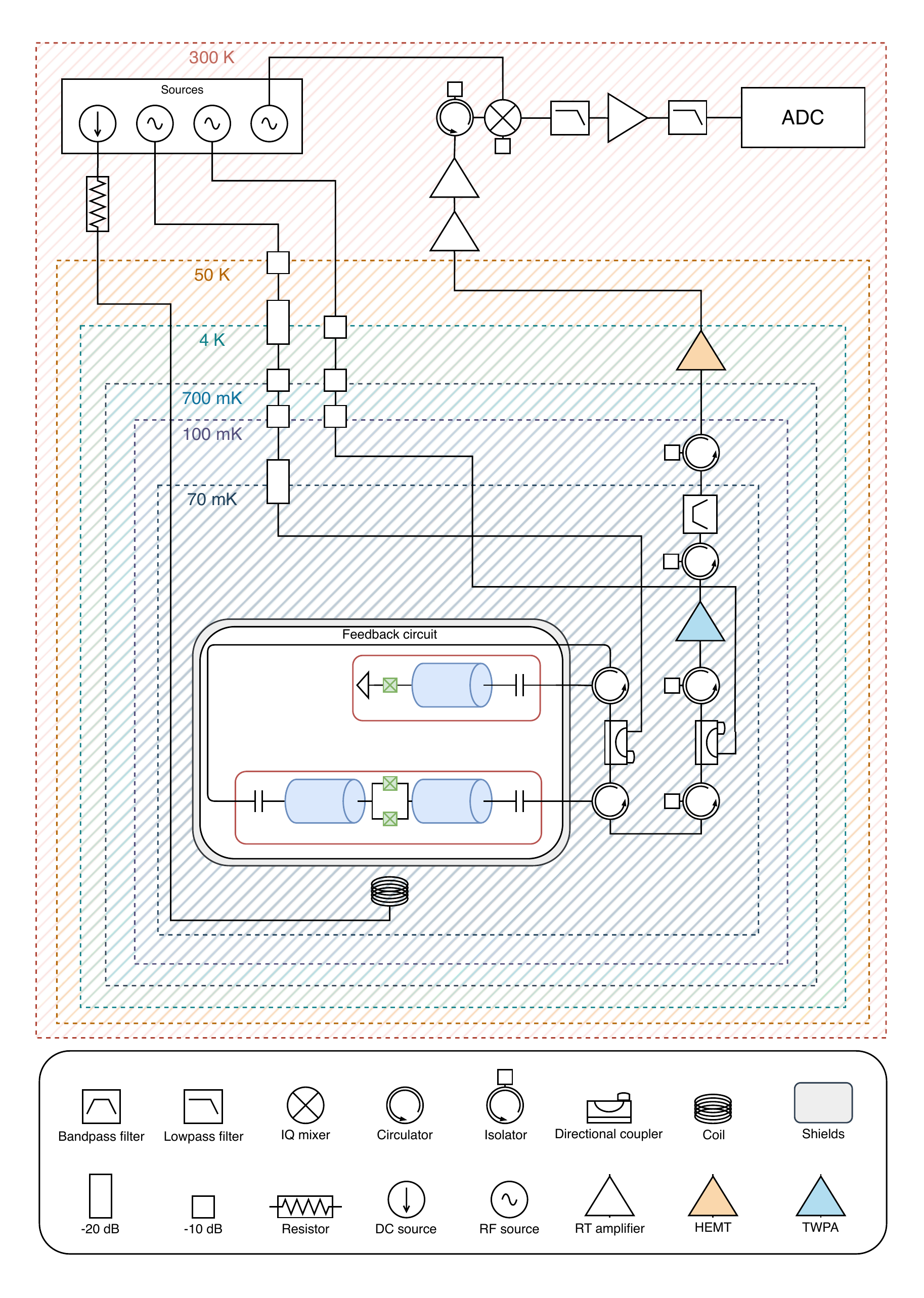}
    \caption{Wiring diagram for the measurement of the coherent feedback circuit. }
    \label{fig:fridge diagram}
\end{figure}
A wiring diagram of the measurement setup is shown in Fig. \ref{fig:fridge diagram}.In the output line, we isolate and amplify the signal using isolators and a TWPA, followed by another isolator, a bandpass filter of 4-8 GHz, another isolator, a HEMT amplifier at 4 K and additional room-temperature amplifiers. 
The signal is then down-converted to 25 MHz using a Marki IQ mixer and isolated and amplified with a low-pass filter and a room-temperature amplifier. Finally, we digitise the signal using an M4i digitiser card,  demodulate the quadratures via digital downconversion to DC and filter high frequencies by a digital FIR filter. We take $4800$ samples at a rate of $125 \times 10^6$. For the energy spectral density measurements, we compute the absolute square of the FFT of the complex IQ data and average for 100 repetitions.

\bibliography{bibliography}

\end{document}